\documentclass[12pt,a4paper]{article}
\usepackage[latin1]{inputenc}
\usepackage{amsmath,bm}
\usepackage{amsmath} 
\usepackage{amsfonts}
\usepackage{amssymb}
\usepackage{float}
\usepackage{cite}
\usepackage{breqn}
\makeatletter
\newcommand{\mathleft}{\@fleqntrue\@mathmargin0pt}
\newcommand{\mathcenter}{\@fleqnfalse}
\makeatother

\usepackage[colorlinks=true,linkcolor=blue,citecolor=red]{hyperref}
\usepackage{graphicx}
\DeclareGraphicsExtensions{.pdf,.png,.jpg}
\usepackage[a4paper, width = 160mm, top = 30mm, bottom = 25mm, 
bindingoffset = 0mm]{geometry}
\def\beq{\begin{equation}}
	\def\eeq{\end{equation}}
\def\bea{\begin{eqnarray}}
	\def\eea{\end{eqnarray}}

\begin{document} 
\begin{center}
	{\Large \bf Quantum tunneling from a new type of generalized Smith-Volterra-Cantor potential}
\vspace{1.3cm}		
		
{\sf  Vibhav Narayan Singh\footnote[1]{e-mail address:\ \ vibhavn.singh13@bhu.ac.in, \ \ vibhav.ecc123@gmail.com},   Mohammad Hasan\footnote[2]{e-mail address:\ \ mhasan@isro.gov.in, \ \ mohammadhasan786@gmail.com}, Mohammad Umar\footnote[3]{e-mail address:\ \ aliphysics110@gmail.com}, Bhabani Prasad Mandal\footnote[4]{e-mail address:\ \ bhabani.mandal@gmail.com, \ \ bhabani@bhu.ac.in  }}
	
\bigskip
{\em $^{1,4}$Department of Physics,
	Banaras Hindu University,
	Varanasi-221005, INDIA \\ 
$^{2}$Indian Space Research Organisation,
Bangalore-560094, INDIA, \\
$^{3}$Optics and Photonics Centre, Indian Institute of Technology, Delhi-110016}

\bigskip	
%
\noindent {\bf Abstract}		
\end{center}
In this paper, we introduce and analyze the Smith-Volterra-Cantor potential of power \( n \), denoted as SVC\(\left(\rho, n\right)\). Bridging the gap between the general Cantor and SVC systems, this novel potential offers a fresh perspective on Cantor-like potential systems within quantum mechanics that unify fractal and non-fractal potentials. Utilizing the Super Periodic Potential (SPP) formalism, we derive the close form expression of the transmission probability \( T_{G}(k) \). Notably, the system exhibits exceptionally sharp transmission resonances, a characteristic that distinguishes it from other quantum systems. Furthermore, the multifaceted transmission attributes of the SVC\(\left(\rho, n\right)\) are found to be critically dependent on both parameters, \( \rho \) and \( n \), offering an intricate interplay that warrants deeper exploration. Our findings highlight a pronounced scaling behavior of reflection probability with \( k \), which is underpinned by analytical derivations.
\medskip
\vspace{1in}
\newpage	
\section{Introduction}
Quantum tunneling, identified in the late 1920s, has emerged as a significant phenomenon that deepened our grasp of quantum mechanics and its myriad complexities \cite{nordheim, gurney}. Its recognition offered a remarkable insight into the probabilistic nature of quantum systems and bolstered the development and validation of quantum theories. This phenomenon, often described as the passage of particles through barriers deemed insurmountable in classical physics, has been pivotal in both theoretical and experimental physics research. From its early conceptualizations to contemporary interpretations, tunneling has maintained a central position in quantum mechanics research, elucidated by several authors \cite{condon, bohm, wigner, esaki}. Driven by its theoretical significance and potential practical applications, extensive investigations have explored how matter waves propagate through various potential distributions \cite{book01, book02}. Such efforts span a diverse array of quantum mechanical domains, including non-Hermitian quantum mechanics (NHQM) \cite{nh01, nh02, nh03, nh04}, space-fractional quantum mechanics (SFQM) \cite{guo, oli, tare, fractal_fractional,hasan2018_tunneling,hasan2020_tunneling}, and quaternionic quantum mechanics (QQM) \cite{sobhani,hassan01,sobhani01, de, davies, de01,hasan2020_quaternionic}. With an array of contributions over the past century, quantum tunneling and the associated analytical calculations of scattering coefficients continue to be a focal point of quantum research \cite{nordheim, gurney, condon, bohm, wigner, esaki, book01, book02 , giaever, burstein, josephson, lauhon,griffiths}.\\
\indent
Fractals, as initially conceptualized by Mandelbrot \cite{mandelbrot}, are intricate structures characterized by their distinct property of self-similarity. This means that their complex patterns repeat at every scale, whether one zooms in or out of the structure. Originating from mathematical constructs, fractals have transcended their theoretical roots to provide insights into natural phenomena, successfully modeling various intricate formations found in nature \cite{mandelbrot, voss, hurd, falconer, berry}. One of the captivating applications of fractals in quantum mechanics is the exploration of fractal potentials. Among these, the Cantor fractal potential stands out as a foundational member of the fractal family. It serves as a prototype for understanding quantum behaviors in fractal systems. The iterative development process of this potential, alongside related potentials like the Smith-Volterra-Cantor (SVC) system, has attracted research attention recently \cite{svc_tunneling, ucp}. A significant advancement in understanding fractal potentials in quantum mechanics comes from the use of the transfer matrix method. Applied to the Cantor fractal potential, this method has uncovered rich details about scattering processes within such systems \cite{svc_tunneling, ucp, cantor_f1, cantor_f2, cantor_f3, cantor_f4, cantor_f5, cantor_f6, cantor_f7, cantor_f8, cantor_f9, cantor_f10}. Furthermore, it has paved the way for the introduction of newer models, notably the fractal Kronig-Penny model \cite{fractal_kroning_penny} and new type of modified Cantor like systems \cite{svc_tunneling, ucp}.\\
\indent
In this article, we introduce an expansive class of Cantor like potential systems that seamlessly bridges the general Cantor (GC) and general SVC (GSVC or SVC($\rho$)) systems. We designate this potential as the Smith-Volterra-Cantor potential of  \( n \) or SVC($\rho, n$). Through the SPP formalism that we previously established \cite{mh_spp}, we provide a rigorous analysis of transmission coefficients from this potential system. The construction methodology for the SVC($\rho, n$) potential of a particular stage \( G \) is reminiscent of the approach employed for the general SVC system for a corresponding stage. Characterizing this potential for stage \( G \) is governed by three pivotal parameters: \( \rho \), \( n \) and \( G \). To elucidate, the Smith-Volterra-Cantor potential of  \( n \) or SVC($\rho, n$) system is developed by excluding a fraction \(\frac{1}{\rho^{G^n}}\) from the length of the potential segment at stage \( G-1 \). This subtraction is orchestrated at the midpoint of each extant segment of the potential at every progressive stage \( G \). It's imperative to note that while \( \rho \) persists as a positive real number greater than 1, \( n \) is firmly entrenched within the realm of real numbers (\( n \in \mathbb{R} \)). When \( n = 0 \), the system aligns conspicuously with the traditional Cantor fractal system. Conversely, when \( n = 1 \), it embodies the attributes of the general SVC($\rho$) potential system \cite{svc_tunneling}. This unveils the fact that the quintessential Cantor and SVC($\rho$) potential systems are enveloped as intrinsic special cases under the overarching canopy of the Smith-Volterra-Cantor potential of power \( n \). Thus, this potential, adeptly blending the intricacies of both fractal and non-fractal systems, stands as an intriguing subject for delving deep into its inherent transmission properties.\\
\indent
We organize the paper as follows: In section \ref{gsvc_section} we introduce the concept of SVC($\rho, n$) potential. Section \ref{spp_svc} explains how SVC($\rho, n$) system of arbitrary stage $G$ is a special case of rectangular SPP of order $G$. In section \ref{svcn_calc} we calculate the close form expression of transmission probability for general SVC($\rho, n$) potential of arbitrary stage $G$ and discuss various graphical results. Further, the saturation of the transmission probability and the scaling law of the reflection coefficient have been discussed in section \ref{saturationn} and section \ref{scaling} respectively. Finally, the paper culminates with the conclusion and discussions presented in section \ref{conclusion}.   
\section{Generalized Smith-Volterra-Cantor potential of power \texorpdfstring{$n$}{n}}
\label{gsvc_section}
\begin{figure}[H]
\centering
\includegraphics[scale=0.4]{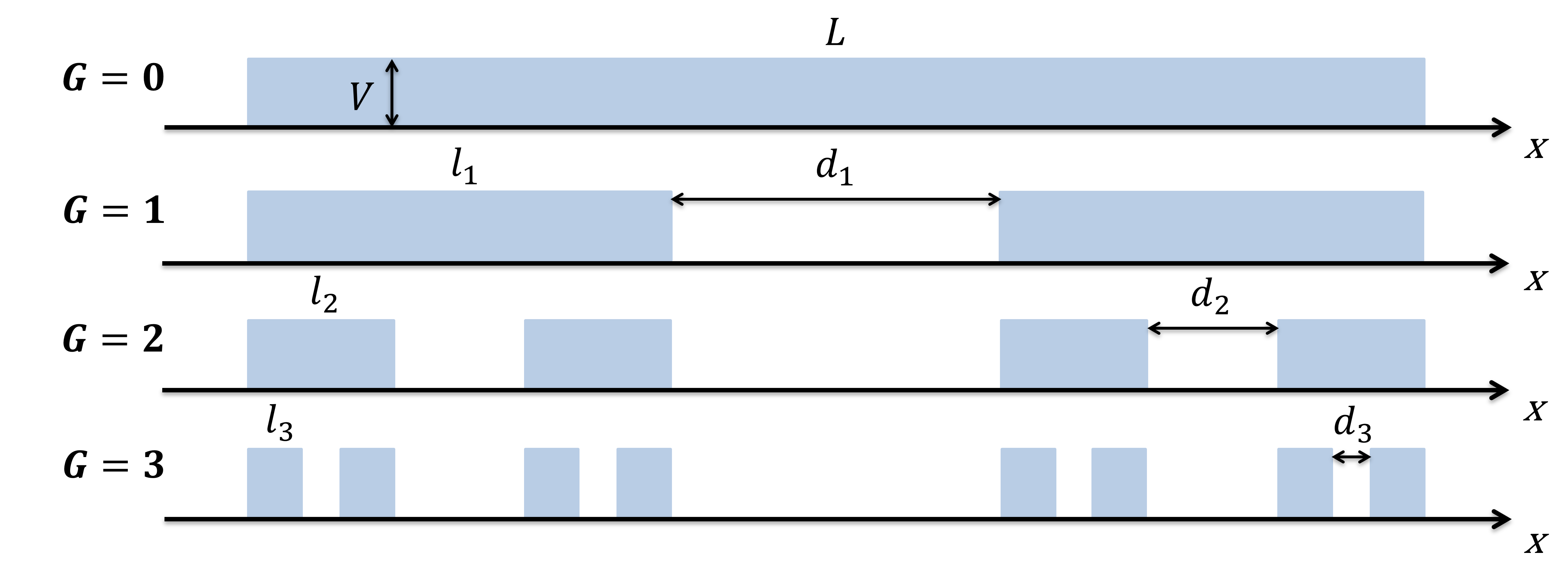}
\caption{\it Graphical representation of the SVC\((\rho)\) and SVC\((\rho, n)\) potential. The white regions demarcate gaps in the potential, with \(V\) symbolizing the potential's height. At every \(G^{th}\) stage, \(l_{G}\) denotes each potential segment's length. For SVC\((\rho)\) potential, the gap \(d_{G}\) equals \(\frac{l_{G-1}}{\rho^G}\), while for SVC\((\rho,n)\), \(d_{G}\) is equivalent to \(\frac{l_{G-1}}{\rho^{G^{n}}}\).}  
\label{svc01}
\end{figure}

This section delves into the generalized Smith-Volterra-Cantor (GSVC) potential of power \(n\), represented as SVC\((\rho, n)\). The traditional SVC potential starts with a rectangular barrier of height \(V\) and length \(L\) at stage \(G=0\). The formation of this potential is based on an iterative process where, at each stage \(G\), a fraction \(\frac{1}{4^{G}}\) of the remaining segment's length is subtracted from the center of the segments at stage \(G-1\). This methodology results in a distinctive fractured pattern that epitomizes the SVC potential. Transitioning to the GSVC potential, the subtraction fraction is adjusted to \(\frac{1}{\rho^{G}}\), where \(\rho > 1\) is a positive real number. This generalization permits a more flexible construction, accommodating patterns beyond the standard SVC potential.  The procedure is initiated at \(G=1\), with a fraction \(\frac{1}{\rho}\) being removed from the center of the length \(L\). Subsequent stages continue this approach: at \(G=2\), an additional \(\frac{1}{\rho^{2}}\) fraction is removed from the two segments remaining post the \(G=1\) phase. As the generations progress, the subtraction fraction from each subsequent segment intensifies. For instance, during the \(G=3\) stage, a segment \(\frac{1}{\rho^{3}}\) is extracted from the quartet of segments left from stage \(G=2\). This process continues for any arbitrary stage $G$ to obtain the GSVC potential of stage $G$ \cite{svc_tunneling}.\\
\indent
In the context of the GSVC potential of power \(n\) denoted as SVC\((\rho, n)\) , the fraction \(\frac{1}{\rho^{G^{n}}}\) is removed at each stage \(G\) instead of \(\frac{1}{\rho^{G}}\), where $n$ is a real number.  This nuanced adjustment has a profound impact on the construction, resulting in a more intricate pattern. With reference to Fig. \ref{svc01}, the SVC\((\rho, n)\) potential is graphically represented upto stage $G=3$, showcasing the systematic removal of fractions at different stages. It is evident that at any stage  \(G\), there exists \(2^{G}\) segments of identical length, denoted as \(l_{G}\). The derivation of \(l_{G}\) can be achieved using the depicted constructs as will be shown later. For $n=1$, SVC\((\rho, n=1)\) = SVC\((\rho)\) and for $n=0$, SVC\((\rho, n=0)\)  = Cantor-$\frac{1}{\rho}$  where Cantor-$\frac{1}{\rho}$ represents general Cantor (GC) potential. For $n=0$ and $\rho=3$, we have standard Cantor-$\frac{1}{3}$ potential system.  Therefore, the SVC\((\rho, n)\) potential encompasses both the Cantor as well as SVC potential system and thus also provides a system through which the transmission features can be studied when a system transits from a fractal to a non-fractal potential system. This unique  SVC\((\rho, n)\) potential system joins a broader class of family of fractal and non-fractal systems into a single unified system. Below we arrive at the expression of $l_{G}$ as follows.
The geometry depicted in Fig. \ref{svc01} illustrates that the potential segment length \(l_{1}\) at stage \(G=1\) is given by 
\begin{equation}
l_{1} = \frac{L}{2}\left(1-\frac{1}{\rho^{1^{n}}}\right).
\end{equation}
For the next stage \(G=2\), the segment length \(l_{2}\) becomes
\begin{equation}
l_{2} = \frac{l_{1}}{2}\left(1-\frac{1}{\rho^{2^{n}}}\right) = \frac{L}{2^{2}}\left(1-\frac{1}{\rho^{1^{n}}}\right)\left(1-\frac{1}{\rho^{2^{n}}}\right).
\end{equation}
Following this pattern,
\begin{equation}
l_{3} = \frac{l_{2}}{2}\left(1-\frac{1}{\rho^{3^{n}}}\right) = \frac{L}{2^{3}} \left(1-\frac{1}{\rho^{1^{n}}}\right)\left(1-\frac{1}{\rho^{2^{n}}}\right)\left(1-\frac{1}{\rho^{3^{n}}}\right),
\end{equation}
and
\begin{equation}
l_{4} = \frac{l_{3}}{2}\left(1-\frac{1}{\rho^{4^n}}\right) = \frac{L}{2^{4}} \left(1-\frac{1}{\rho^{1^{n}}}\right)\left(1-\frac{1}{\rho^{2^{n}}}\right)\left(1-\frac{1}{\rho^{3^{n}}}\right)\left(1-\frac{1}{\rho^{4^n}}\right).
\end{equation}
By extending these steps, the segment length \(l_{G}\) for any arbitrary \(G^{th}\) stage SVC\((\rho, n)\) potential is derived as
\begin{equation}
l_{G} = \frac{L}{2^{G}}\prod_{j=1}^{G}\left(1-\frac{1}{\rho^{j^{n}}}\right). 
\label{l_GG}
\end{equation}
This product is not represented in a known form unless \(n=1\). In that special case, where the potential is defined as SVC\((\rho)\), \(l_{G}\) is described by \cite{svc_tunneling}
\begin{equation}
l_{G} = \frac{L}{2^{G}} \times q \left ( \frac{1}{\rho}; \frac{1}{\rho} \right)_{G} .
\label{l_G_qp}
\end{equation}
Here, the symbol \(q\) denotes the \(q\)-Pochhammer symbol \cite{abramowitz}, which is defined as
\begin{equation}
q(x;y)_{p}=\prod_{j=0}^{p-1}(1-x.y^{j})=(1-x)(1-x.y)(1- x. y^{2})\ldots(1-x.y^{p-1}).
\label{qp}
\end{equation} 
\section{SVC\texorpdfstring{$(\rho, n)$}{(rho, n)} potential as a special case of super periodic rectangular potential}
\label{spp_svc}
\begin{figure}[H]
\begin{center}
\includegraphics[scale=0.51]{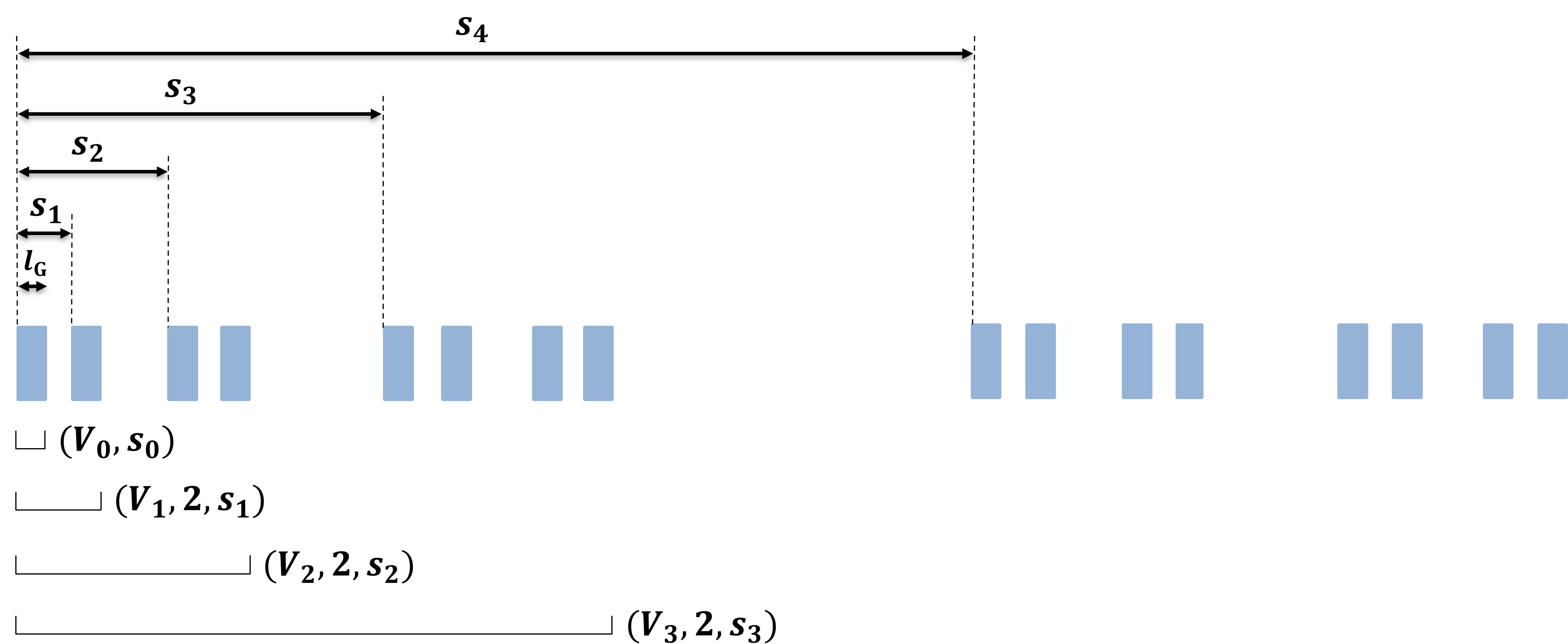}
\caption{\it Construction of the SVC($\rho, n$) potential of order $4$ (or stage $G =4$) through the SPP formalism. Here, $s_{1}$, $s_{2}$, $s_{3}$ and $s_{4}$ are the periodic distance when the unit cell potentials $V_{0}$, $V_{1}$, $V_{2}$ and $V_{3}$ are periodically repeated respectively with periodicity $2$.} 
\label{svc02}
\end{center}
\end{figure}

In the context of the concept of super periodic potential (SPP) introduced in reference \cite{mh_spp}, we briefly expound upon this concept for the sake of comprehensiveness within the scope of this paper. Commencing with a unit cell potential denoted as $V$, we construct a periodic system by repetitively placing the unit cell system in a consecutive manner, with a specified finite repetition count denoted as $N_{1}$, and at regularly spaced intervals denoted as $s_{1}$. Here, $s_{1}$ signifies the distance between the starting positions of two consecutive unit cells. The resulting periodic potential is denoted as $V_{1}=(V, N_{1},s_{1})$. Subsequently, call the system $V_{1}$ as a new unit cell and iterate this periodic repetition process by introducing another repetition count denoted as $N_{2}$ and another regular spacing interval denoted as $s_{2}$, resulting in the formation of the system $V_{2}=(V_{1}, N_{2},s_{2})$. Now, call the system $V_{2}$ as a new unit cell and iterate this periodic repetition $N_{3}$ times at consecutive distances $s_{3}$ to obtain another unit cell system $V_{3}=(V_{2}, N_{3},s_{3})$ which is further periodically repeated in a similar manner to obtain $V_{4}=(V_{3}, N_{4},s_{4})$. This sequence of periodic repetitions can extend to an arbitrary finite number of iterations denoted as $g$, resulting in the establishment of the super periodic potential system of order $g$, designated as $V_{g}=(V_{g-1}, N_{g}, s_{g})$. This hierarchical approach allows for the generation of increasingly complex SPP structures through successive iterations of unit cell repetition and spacing adjustments.\\
\indent
Next, we demonstrate that the potential configuration SVC$(\rho, n)$ of stage $G$ is, in fact, a specific instance of a rectangular SPP of order $G$. As previously discussed, the SVC$(\rho, n)$ structure of stage $G$ is created by removing a fraction $\frac{1}{\rho^{G^{n}}}$ from the center of each potential segment at stage $G$. So, stage $G=1$ results in two separate potentials of equal length $l_{1}$, which inherently form a periodic system with $N_{1}=2$ and a periodic distance of $s_{1}=l_{1}+d_{1}$, where $d_{1}$ represents the length between the end of the first potential segment and the start of the second potential segment (as shown in Fig. \ref{svc01}). The construction of the SVC$(\rho,n)$ potential system follows a sequential process in which a specific length as discussed in the previous section is systematically excised from the central portion of the potential segments at each stage $G$. This executed process ensures that, at each stage, the resulting potential segments maintain the same length. This implies that the SVC$(\rho,n)$ potential is an special case of SPP with $N_{p}=2$ and with some $s_{p}$, $p=1,2,3,.....,g$. We initiate this concept from a unit cell rectangular potential barrier, and the construction of the SVC$(\rho, n)$ system of stage $G=4$ is visually represented in Fig. \ref{svc02}. Notably, it is worth noting that the total number of super periodic repetition $g$ is equal to the stage $G$ of the SVC$(\rho, n)$ potential and therefore $g=G$. This shows that SVC$(\rho, n)$ potential of stage $G$ and height $V$ is the special case of rectangular SPP of order $G$ in which unit cell potential is a rectangular potential of height $V$ and width $l_{G}$. Now, let's calculate the values of $s_{p}$ which is as follows:
\\
\indent
From the definition of the SVC$(\rho, n)$ potential system and its construction through the SPP concept as shown in Fig. \ref{svc02}, and also using the geometry of GSVC potential illustrated in Fig. \ref{svc01} it is noted that,  
\begin{equation}
s_{1} =l_{G}+\frac{l_{G-1}}{\rho^{G^{n}}}.  \nonumber
\end{equation}
Similarly,
\begin{equation}
s_{2} = l_{G-1}+\frac{l_{G-2}}{\rho^{(G-1)^{n}}},   \nonumber
\end{equation}
 \begin{equation} 
s_{3} = l_{G-2}+\frac{l_{G-3}}{\rho^{(G-2)^{n}}},   \nonumber
\end{equation}
\begin{equation} 
s_{4} = l_{G-3}+\frac{l_{G-4}}{\rho^{(G-3)^{n}}}.   \nonumber
\end{equation}
\noindent
Therefore for a general $s_{p}$ we have the following expression,
\begin{equation}
s_{p} =  l_{G+1-p}+\frac{l_{G-p}}{\rho^{(G+1- p)^{n}}}.
\end{equation}
Using Eq. (\ref{l_GG}) in above expression, we obtain the expression of $s_{p}$ as,
\begin{equation}
s_{p}=\frac{L}{2^{G+1-p}} \left( 1+ \frac{1}{\rho^{(G+1-p)^{n}}} \right)  \prod_{j=1}^{G-p} \left( 1-\frac{1}{\rho^{j^{n}}} \right)  .
\label{sp_svcn1}
\end{equation}
\noindent
Further, Eq. (\ref{sp_svcn1}) can also be recognized as,
\begin{equation}
s_{p}= \frac{1}{2} l_{G-p} \left[ 1+ \frac{1}{\rho^{(G+1-p)^{n}}} \right ].
\label{sp_svcn2}	
\end{equation}	
Thus with the knowledge of $s_{p}$, $p=1,2,...G$ and starting from a rectangular barrier of width $l_{G}$ and height $V$ $(\equiv V_{0})$, the $G^{th}$ stage SVC$(\rho, n)$ system can be obtained through the following sequence of operation: $V_{1}= (V_{0},2,s_{1})$, then obtain $V_{2}= (V_{1},2,s_{2})$, then $V_{3}= (V_{2},2,s_{3})$ $,.....$, $V_{G}= (V_{G-1},2,s_{G})$.  The generated system $V_{G}$ will be our $G^{th}$ stage SVC$(\rho, n)$ system. In the next section, we calculate the tunneling probability from the SVC$(\rho, n)$ potential system.
\section{Tunneling probability from SVC\texorpdfstring{$(\rho, n)$}{(rho, n)} potential}
\label{svcn_calc}	
Reference \cite{mh_spp} provides a detailed explanation of a method for obtaining the transfer matrix and corresponding transmission coefficient of a super periodic potential (SPP) with any given order $G$ when the transfer matrix $M(k)$ of unit cell potential is known.\\
The transfer matrix for a unit cell potential is
\begin{equation}
M (k)= \begin{pmatrix}   M_{11} (k) & M_{12} (k) \\ M_{21} (k) & M_{22} (k)  \end{pmatrix}. 
\end{equation}\\
Then the tunneling coefficient for SPP of order $G$ is expressed through
\begin{equation}
T(N_{1},N_{2},.....,N_{G})=\frac{1}{1+\left[|M_{12}|U_{N_{1}-1}(\Omega_{1})U_{N_{2}-1}(\Omega_{2})U_{N_{3}-1}(\Omega_{3})........U_{N_{G}-1}(\Omega_{G})\right]^{2}}. 
\label{t_spp}
\end{equation}
In the above equation,  $U_{N}(\Omega)$ is the Chebychev polynomial of the second kind \cite{abramowitz} and various $\Omega_{q}$s appearing in the above equation are the Bloch phase of the corresponding fully developed periodic system. For the special case when $N_{j}=2$ and the unit cell transfer matrix is a rectangular barrier of height $V$ and width $l_{G}$, we have transmission coefficient for $G^{th}$ stage SVC$(\rho,n)$ potential as,
\begin{equation}
T_{G}(k)=\frac{1}{1+4^{G}\varepsilon_{-}^{2}\sin^{2}{(\kappa l_{G})} \prod_{q=1} ^{G} \Omega_{q}^{2}}.
\label{T_svc}
\end{equation}
In the above we have used $U_{0}(h)=1$, $U_{1}(h)=2h$ and $M_{12}$ element of the transfer matrix of rectangular barrier. The elements of the transfer matrix of the rectangular barrier are given by \cite{griffiths},
\begin{subequations}
\begin{equation}
M_{11}=(\cos{\kappa l_{G}}-i\varepsilon_{+} \sin{\kappa l_{G}})e^{ikl_{G}},
\end{equation}
\begin{equation}	M_{12}=i\varepsilon_{-} \sin{\kappa l_{G}},
\end{equation}
\begin{equation}
M_{21}=-i\varepsilon_{-} \sin{\kappa b},
\end{equation}
\begin{equation}
M_{22}=(\cos{\kappa b}+i\varepsilon_{+} \sin{\kappa l_{G}})e^{-ikl_{G}}.
\end{equation}
\end{subequations}
Where,
\begin{equation}
\varepsilon_{\pm}=\frac{\tau \pm \tau^{-1}}{2}, \ \  \ \ \tau=\frac{k}{\kappa},
\nonumber
\label{epsilon_plus_minus}
\end{equation}
and 
\begin{equation}
k=\frac{\sqrt{2mE}}{\hbar}, \ \  \ \ \kappa=\frac{\sqrt{2m(E-V)}}{\hbar}.\nonumber 
\end{equation} 
The functional form of $\Omega_{j}$ for SVC($\rho,n$) and SVC($\rho$) case is same and is given below \cite{svc_tunneling},
\begin{equation}
\Omega_{q}(k) = 2^{q-1}\vert M_{22} \vert\cos\big\{\theta-k\gamma_{1}(q)\big\} \prod_{p=1}^{q-1} \Omega_{p} - \sum_{r=1}^{q-1}\left\{2^{q-r-1}\cos \big \{k\gamma_{2}(q,r)\big \} \prod_{p=r+1}^{q-1}\Omega_{p}\right\} .
\label{zeta_n1}
\end{equation}
In the above, 
\begin{equation}
\gamma_{1}(q) = \left\{\sum_{p = 1}^{q-1}s_{p}\right\}-s_{q},
\end{equation}
and 
\begin{equation}
\gamma_{2}(q,r) = \gamma_{1}(q)-\gamma_{2}(r).
\label{chi2}
\end{equation}
Eq. (\ref{T_svc}) along with Eq. (\ref{zeta_n1}) completes the calculation for the transmission probability from SVC($\rho,n$) potential provided $l_{G}$ and various $s_{p}, p=1,2,3,..,G$ are calculated from Eq. (\ref{l_GG}) and (\ref{sp_svcn1}) respectively for a given $n$. Further, from the geometry of Fig. \ref{svc01} and \ref{svc02} it can be easily shown that, 
\begin{equation}
    \gamma_{1}(q) = -(l_{G}+d_{G-q+1}),
\end{equation}
which implies that $\gamma_{1}(j)$ is always negative. Using above equation in Eq. (\ref{chi2}), we get
\begin{equation}
    \gamma_{2}(q, r) = d_{G-r+1}-d_{G-q+1}.
\end{equation}
It can be seen in Fig. \ref{svc01} that for $x>y$, $d_{x}<d_{y}$. Also for $r<q$, $G-r+1>G-q+1$. Hence, it is clear that $d_{G-r+1}<d_{G-q+1}$. Therefore, for $r<q$, $\gamma_{2}(q, r) < 0$.
\begin{figure}[H]
\begin{center}
\includegraphics[scale=0.09]{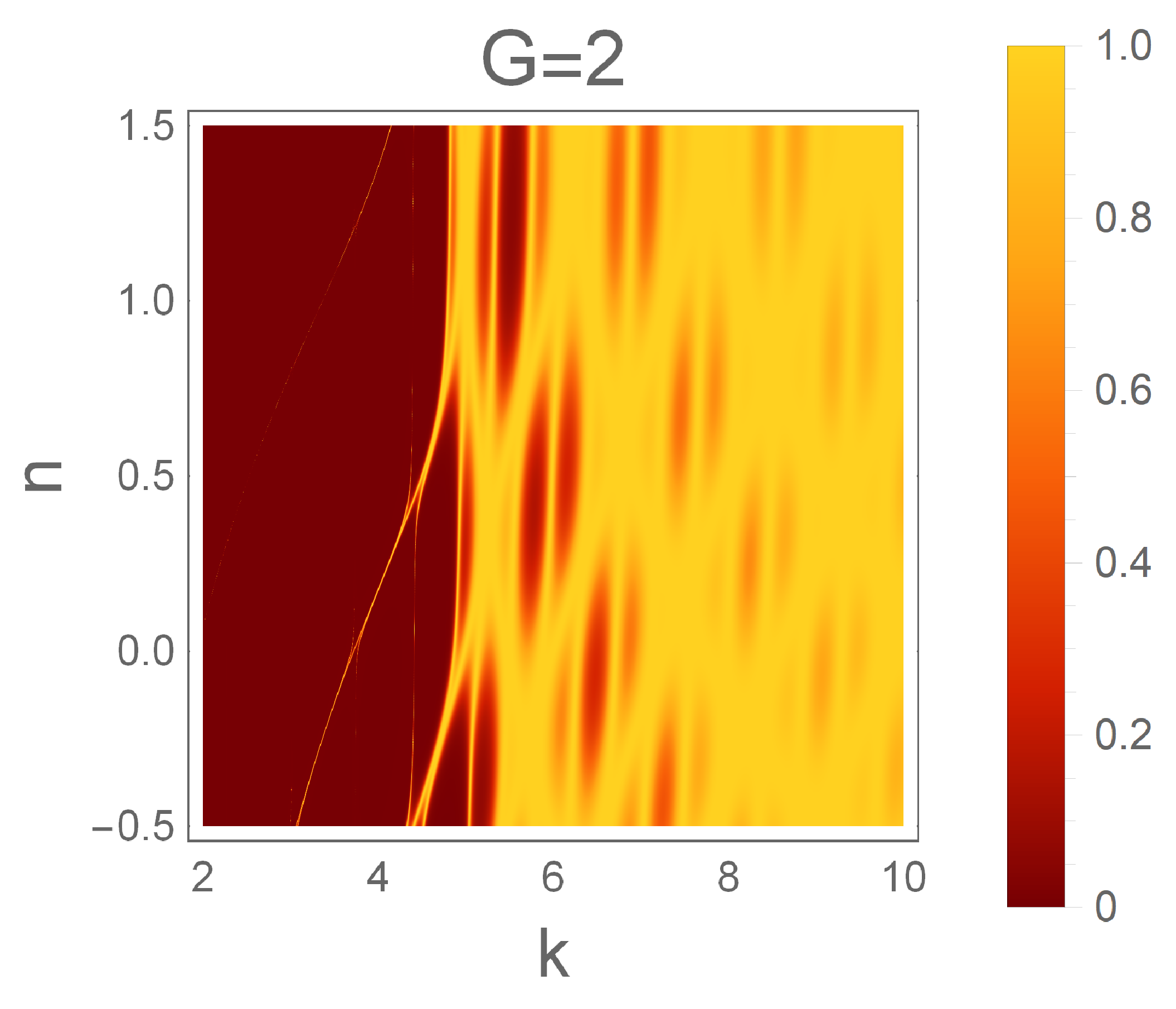} a \ \includegraphics[scale=0.09]{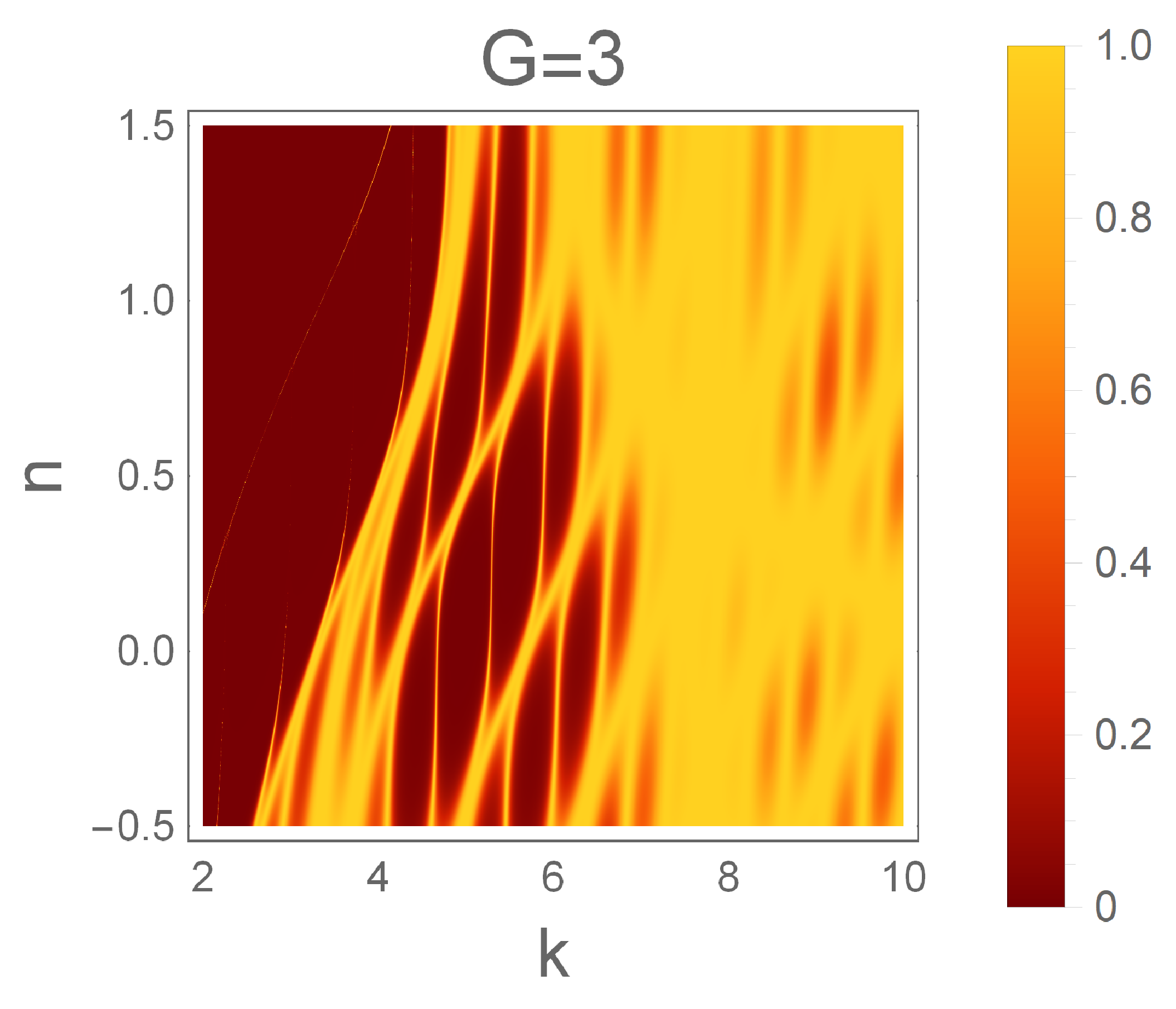} b \\
\includegraphics[scale=0.09]{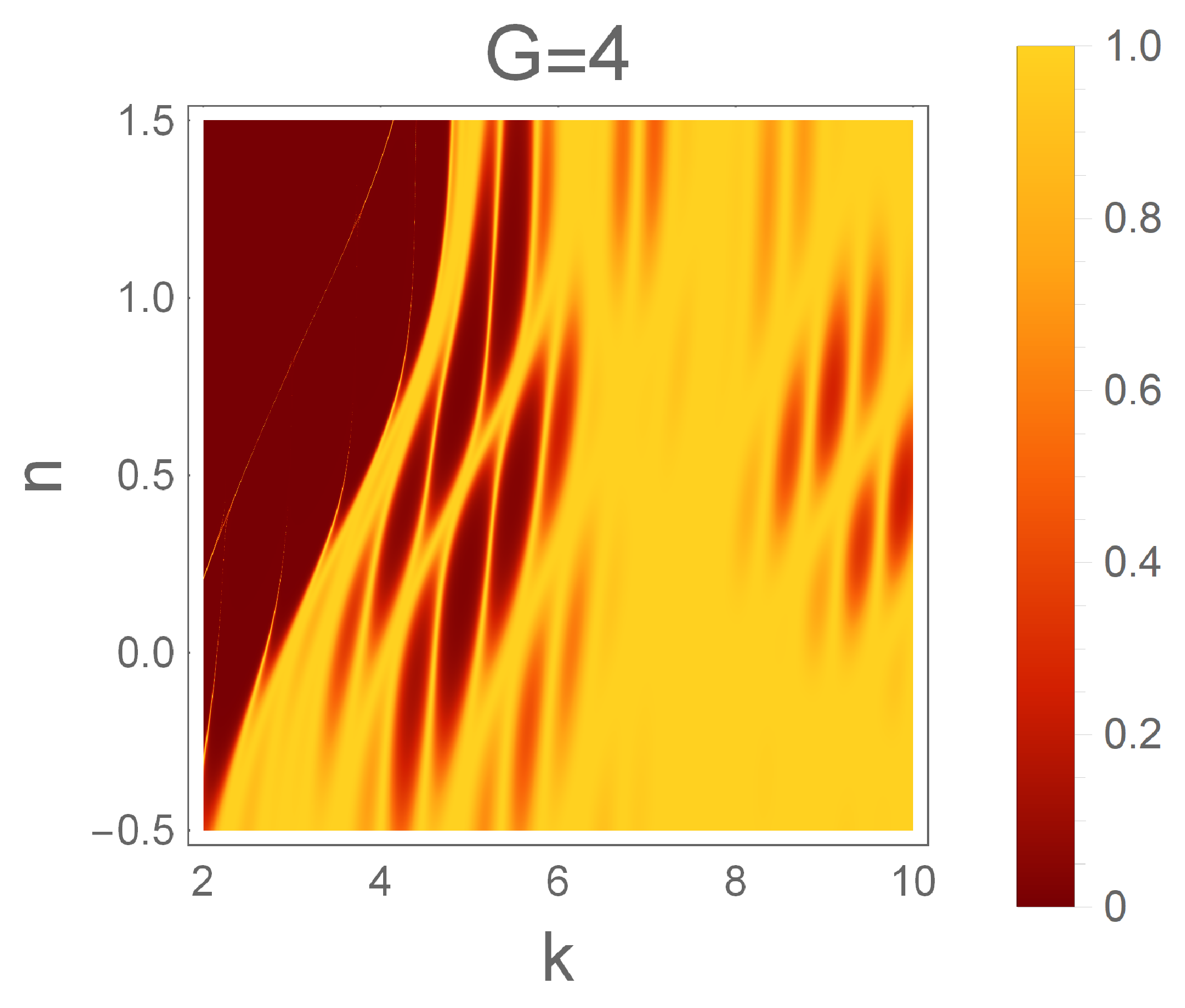} c \ \includegraphics[scale=0.28]{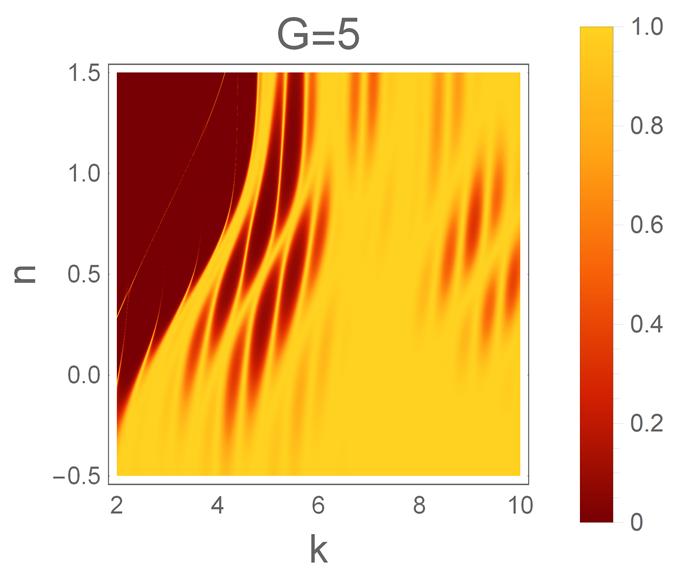} d  
\caption{\it The above figures present density plots depicting the transmission probability $T$ for different stages $G$ of SVC($\rho,n$) potential, where the potential height is set to $V=10$, and the total spatial span is $L=10$. Also, $\rho$ is chosen to be equal to the exponential constant $e$.}  
\label{density_plots_svc}
\end{center}
\end{figure}
\indent
Fig. \ref{density_plots_svc} offers a detailed visualization through density plots of transmission probabilities across multiple potential stages denoted by $G$. These plots, characterized by parameters $V=10$, $L=10$, and $\rho=e$ (the exponential constant), illustrate the intricate behavior of quantum tunneling through the second, third, fourth and fifth stage of SVC($\rho, n$) potential. One of the most salient features observed across the spectrum is the presence of extraordinarily sharp transmission resonances for lower $k$ values. These resonances manifest as sharp peaks where $T=1$, indicating the locus of perfect transmission. An evident observation is the systematic shift in patterns as $G$ progresses. Initial stages portray clearer, less convoluted resonance patterns, which become increasingly intricate with the escalation of $G$. This shift can be likened to the increasing complexity of the quantum dynamics at play, as more potential stages are introduced. 
\begin{figure}[H]
\begin{center}
\includegraphics[scale=0.168]{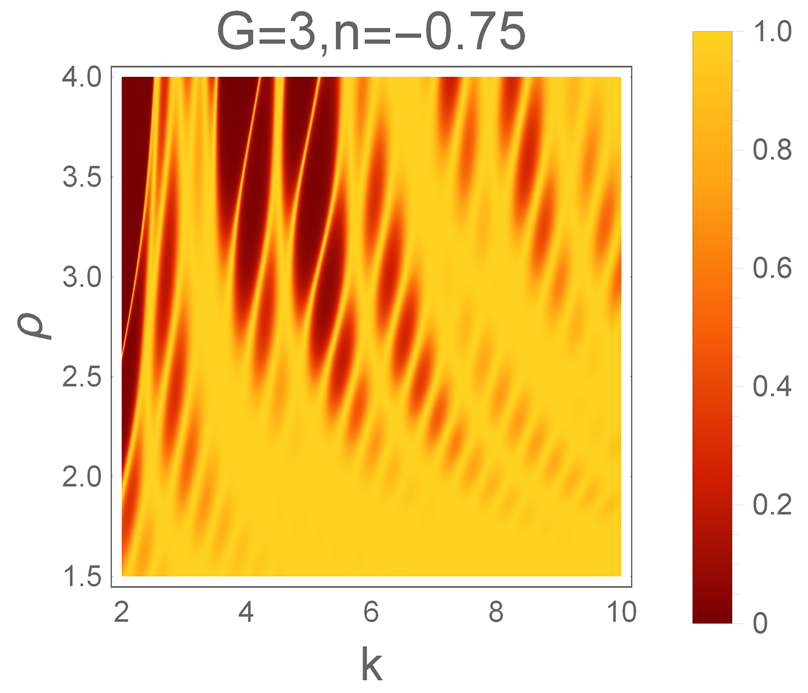}	a		
\includegraphics[scale=0.168]{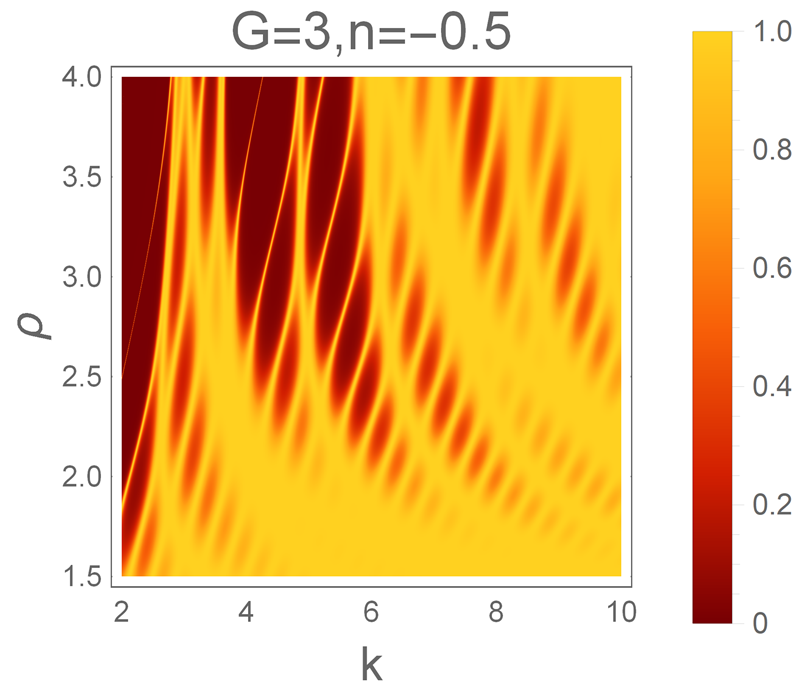} b   
\includegraphics[scale=0.168]{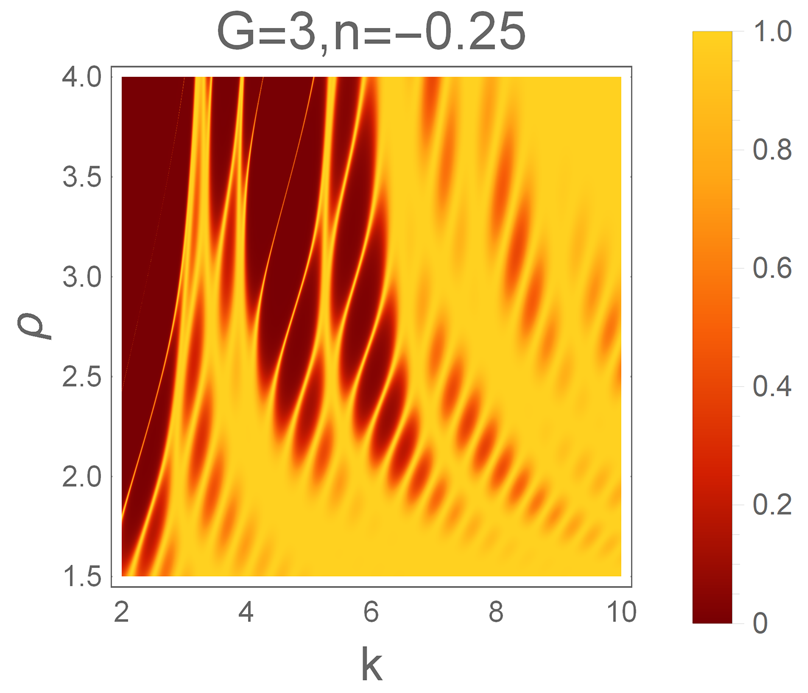} c \\  	
\includegraphics[scale=0.064]{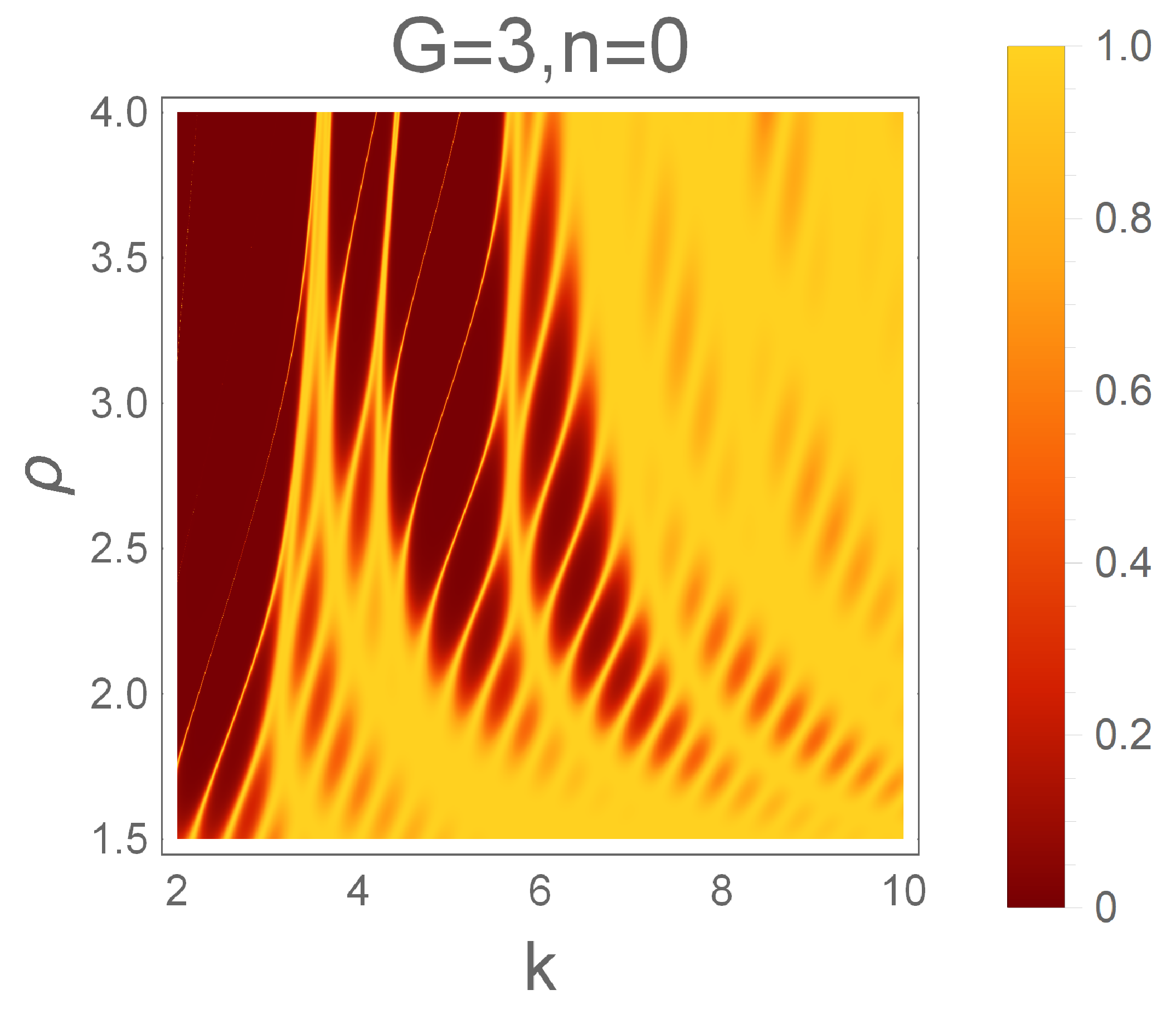} d 
\includegraphics[scale=0.064]{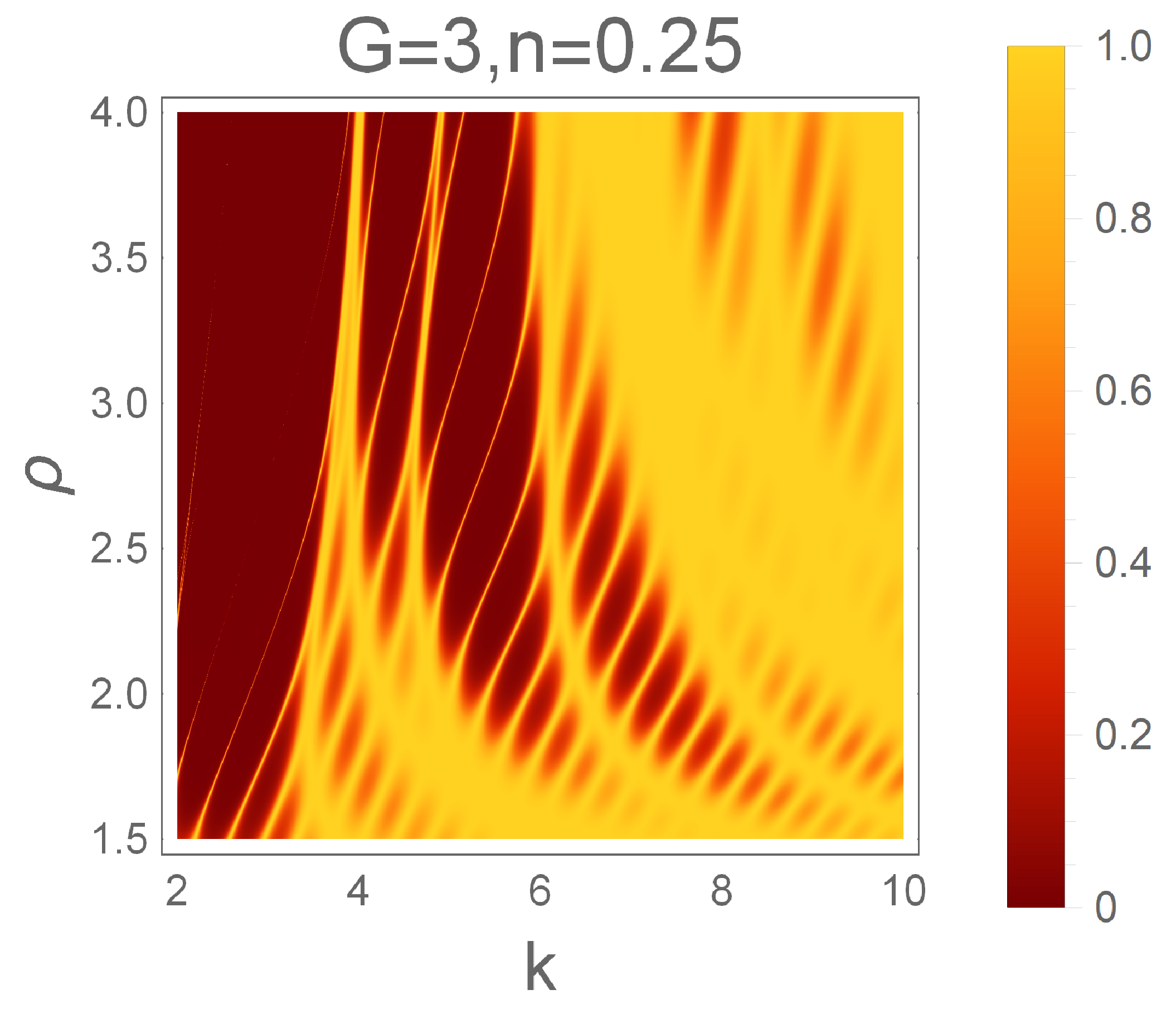} e 
\includegraphics[scale=0.064]{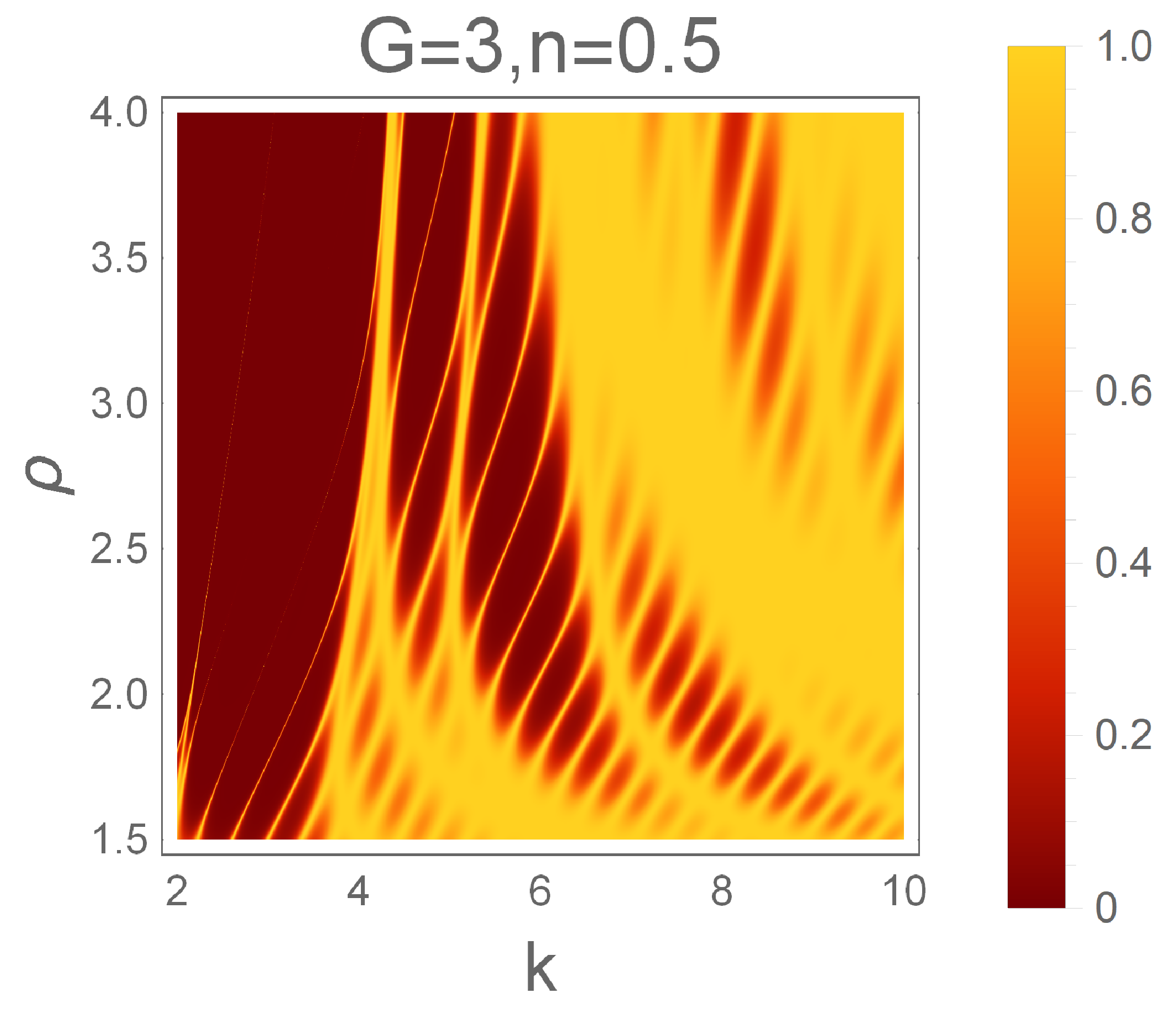} f \\ 
\includegraphics[scale=0.169]{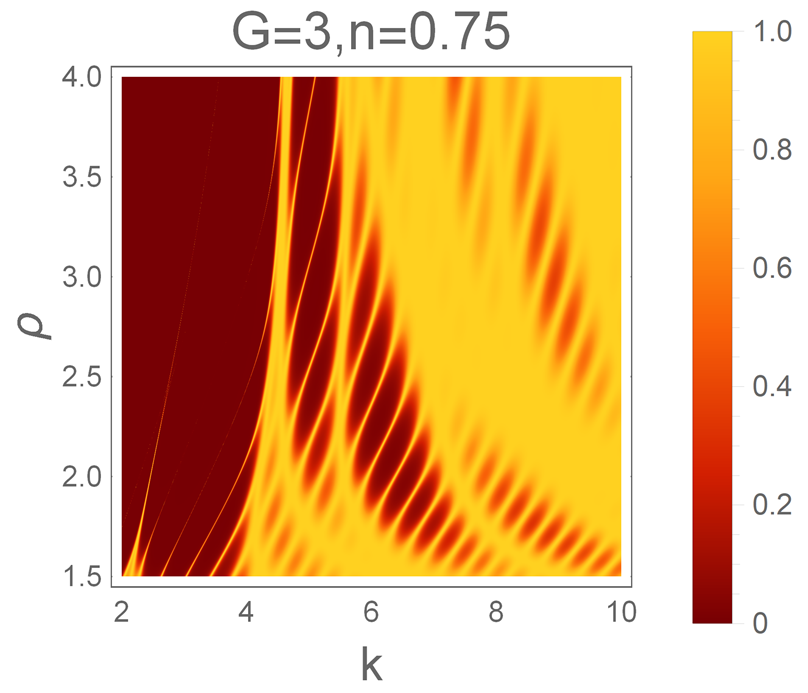} g
\includegraphics[scale=0.17]{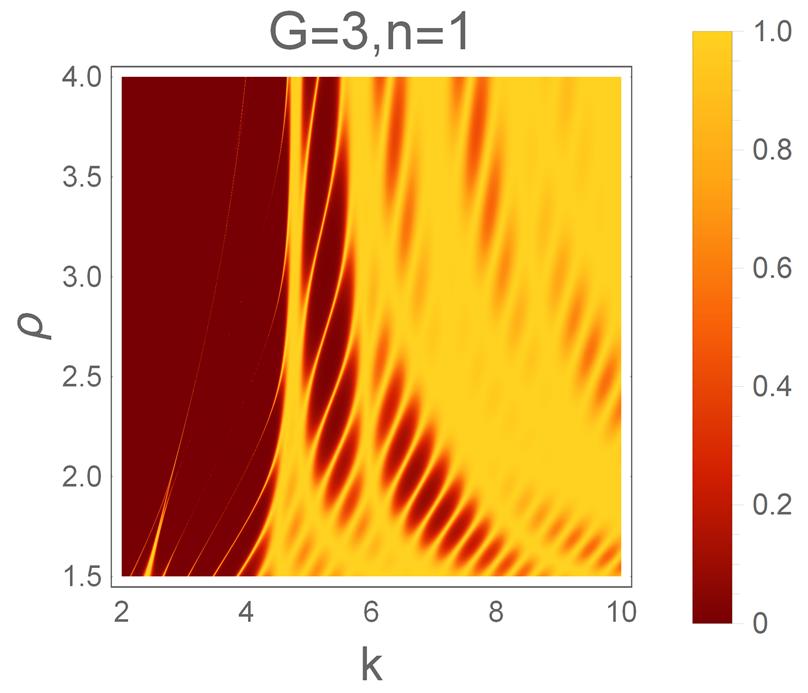} h  
\includegraphics[scale=0.064]{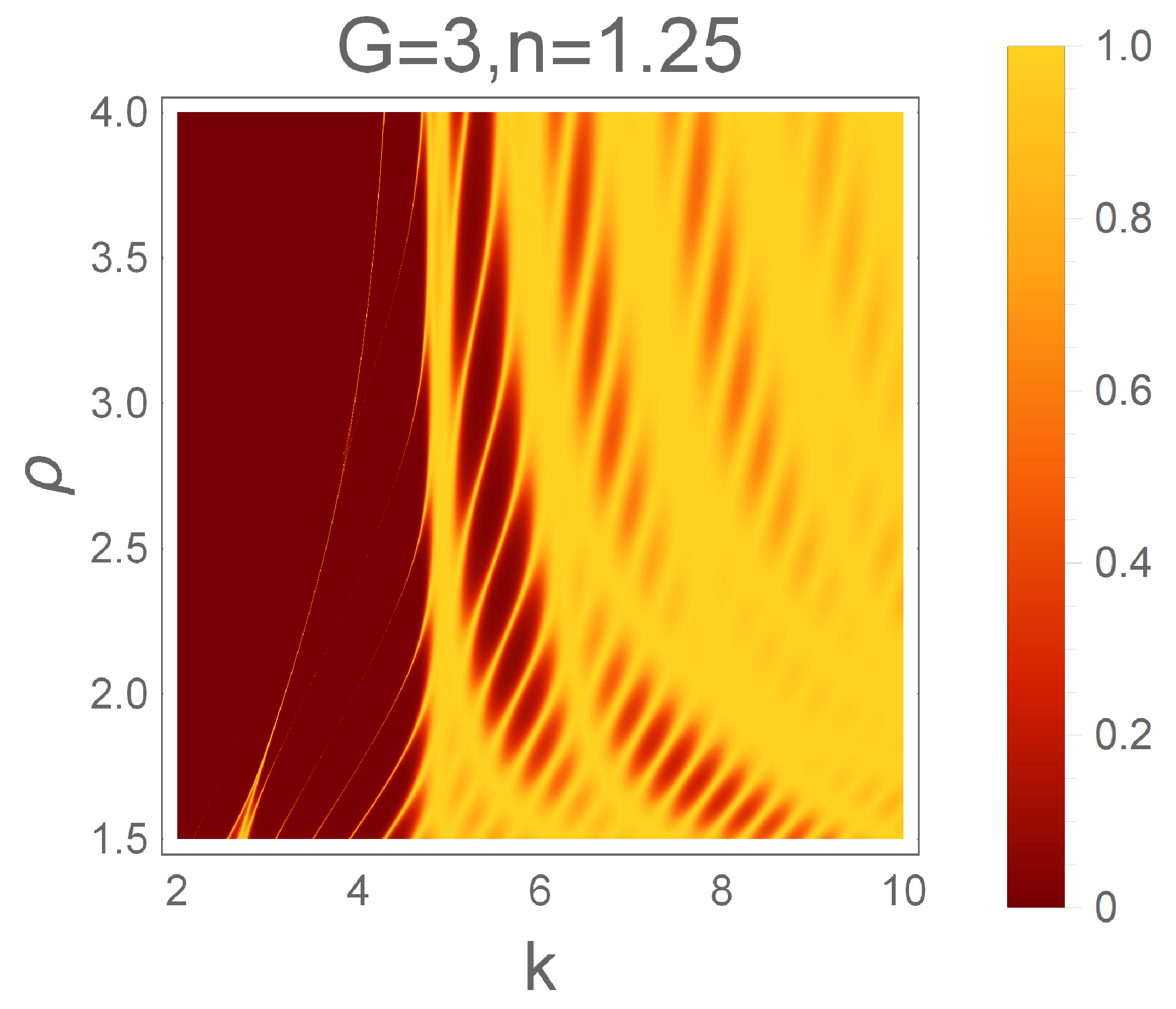} i \\   
\includegraphics[scale=0.064]{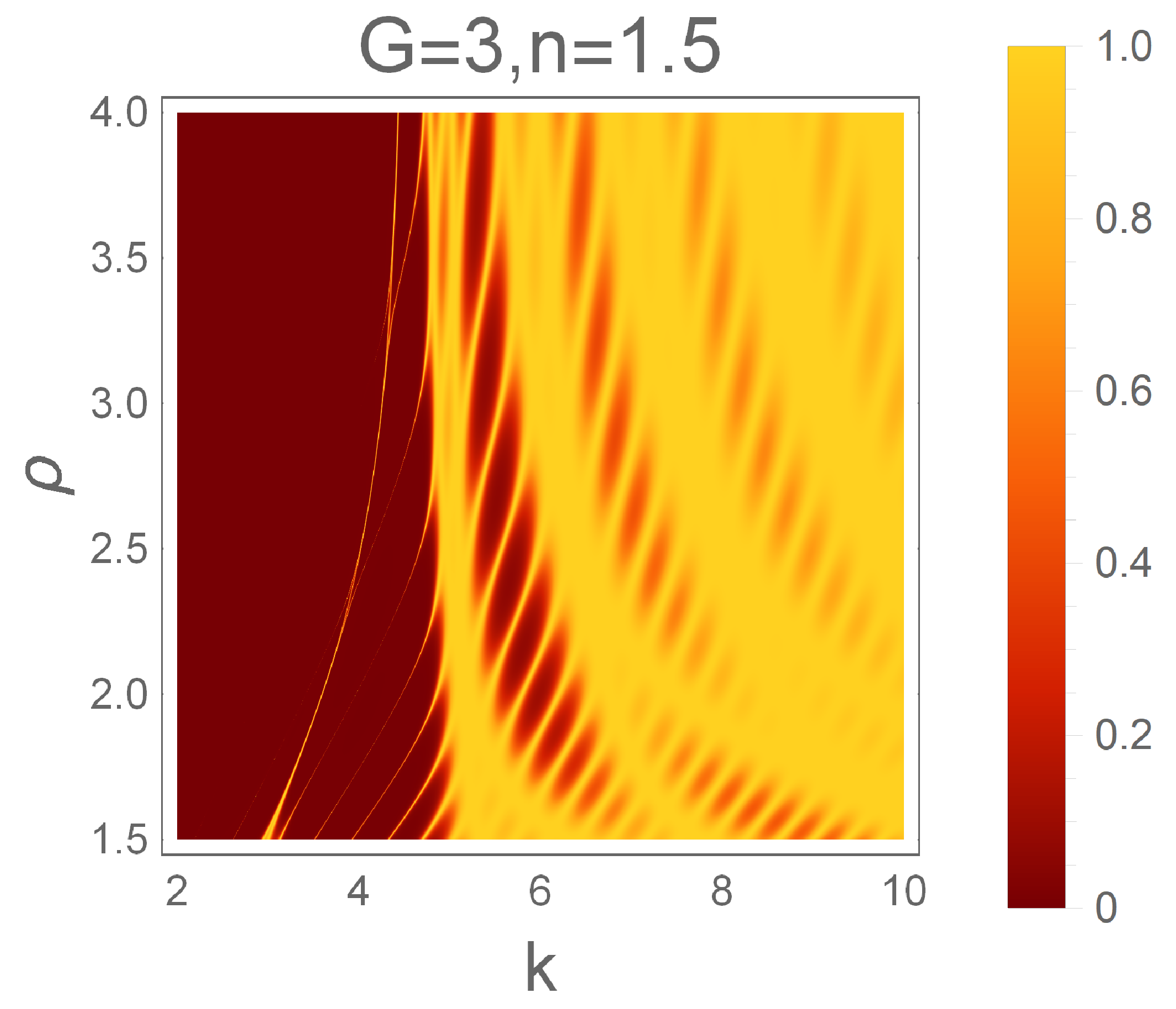} j
\includegraphics[scale=0.064]{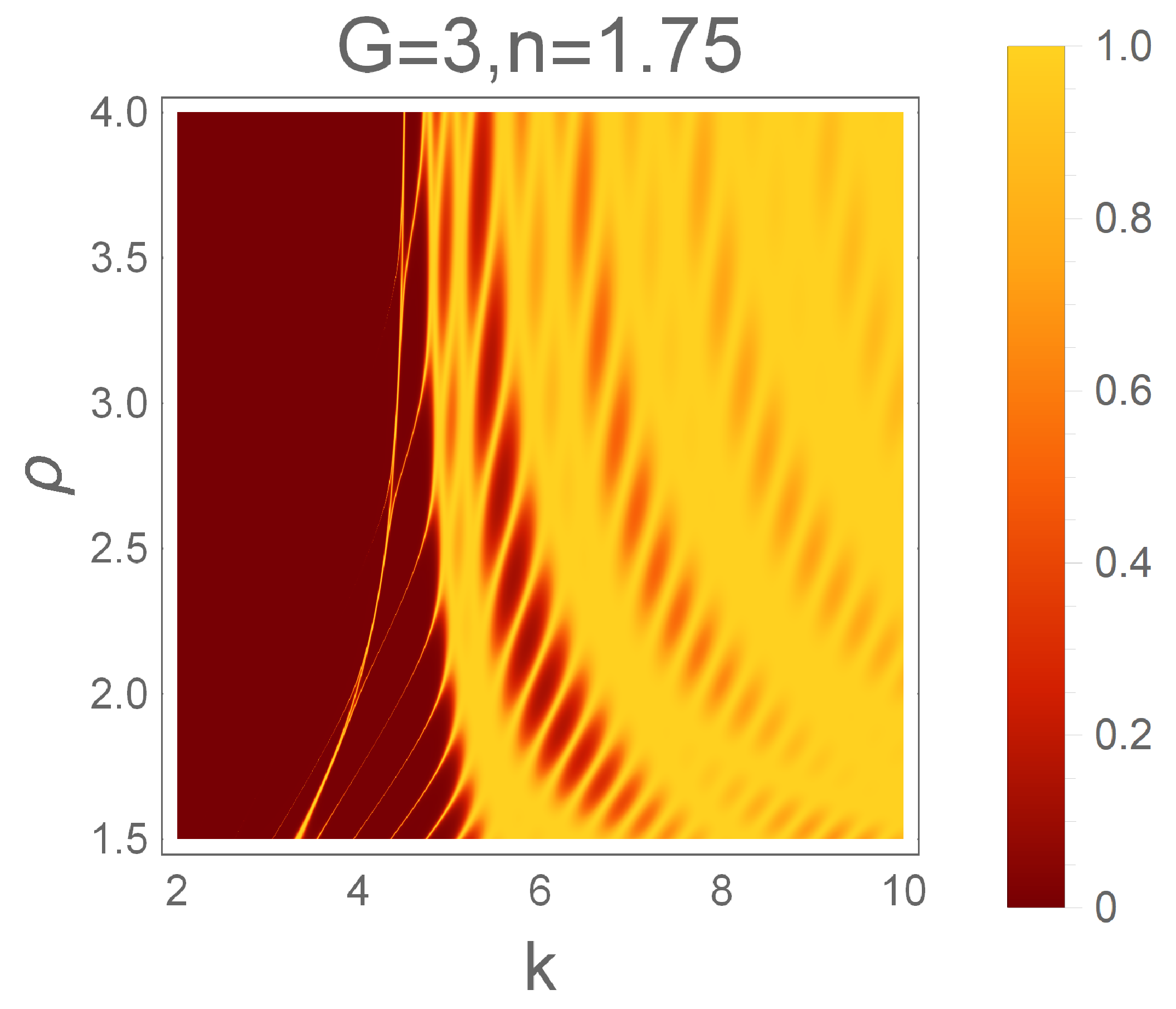} k 
\includegraphics[scale=0.065]{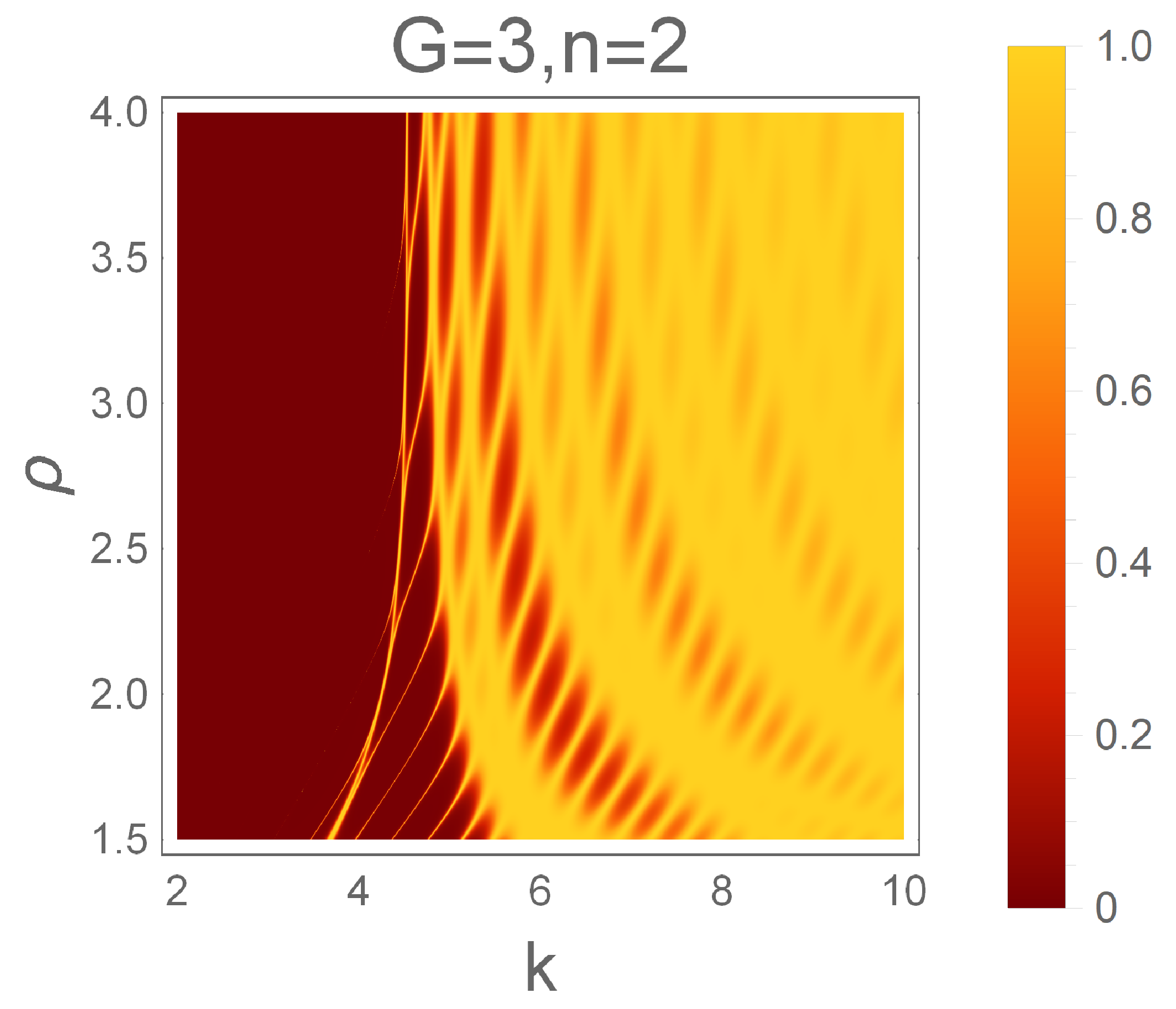} l 
\caption{\it Density plots of tunneling probability $T$ in $\rho-k$ plane for SVC($\rho,n$) potential of different values of $n$ for stage $G=3$. The other potential parameters are $V=10$ and $L=10$. It is seen from these plots that sharper features of transmission resonances occur for lower $k$ values as $n$ increases.}  
\label{dp_nk_g3}
\end{center}
\end{figure}  
\begin{figure}[H]
\begin{center}
\includegraphics[scale=0.065]{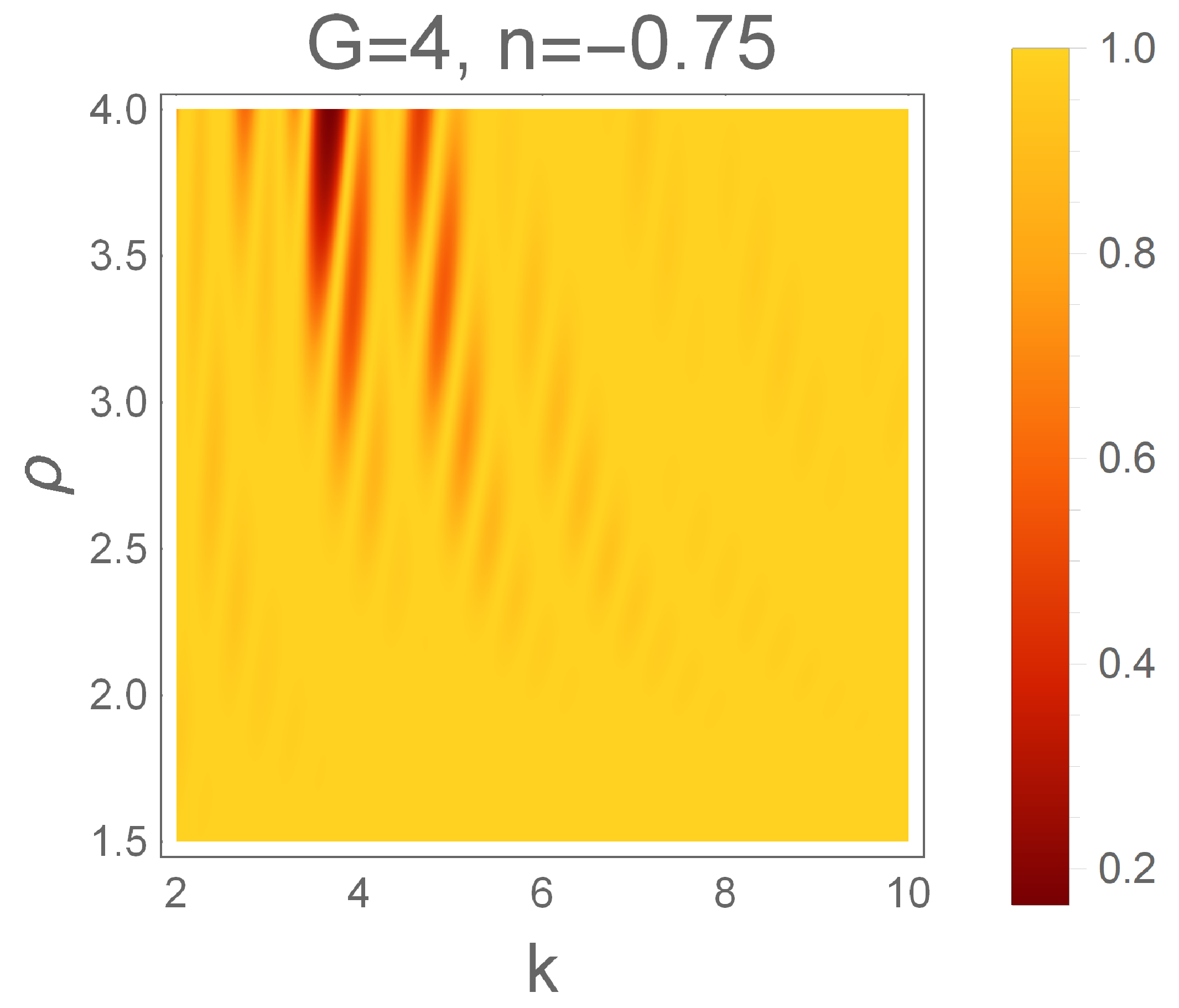} a		
\includegraphics[scale=0.065]{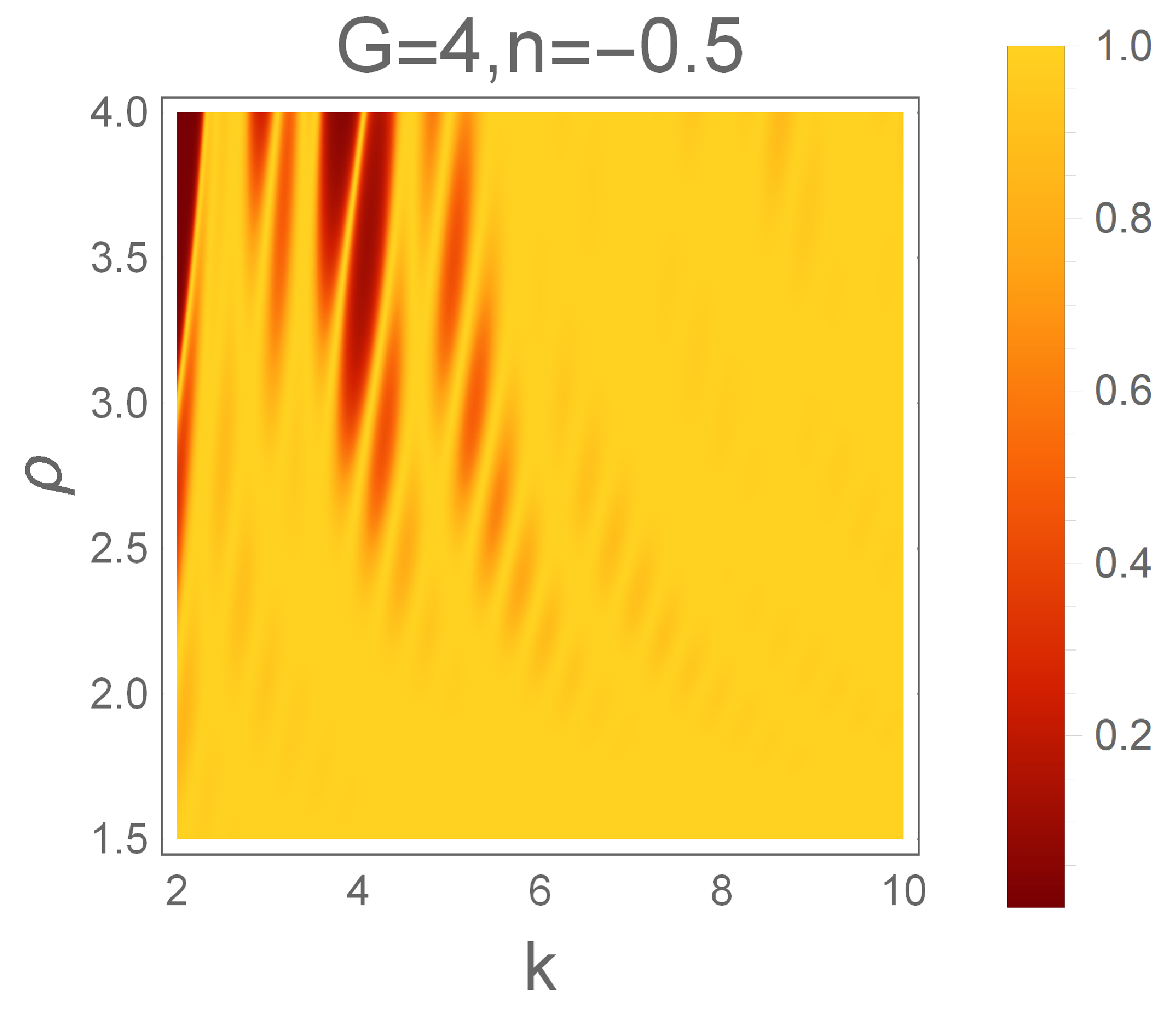} b   
\includegraphics[scale=0.065]{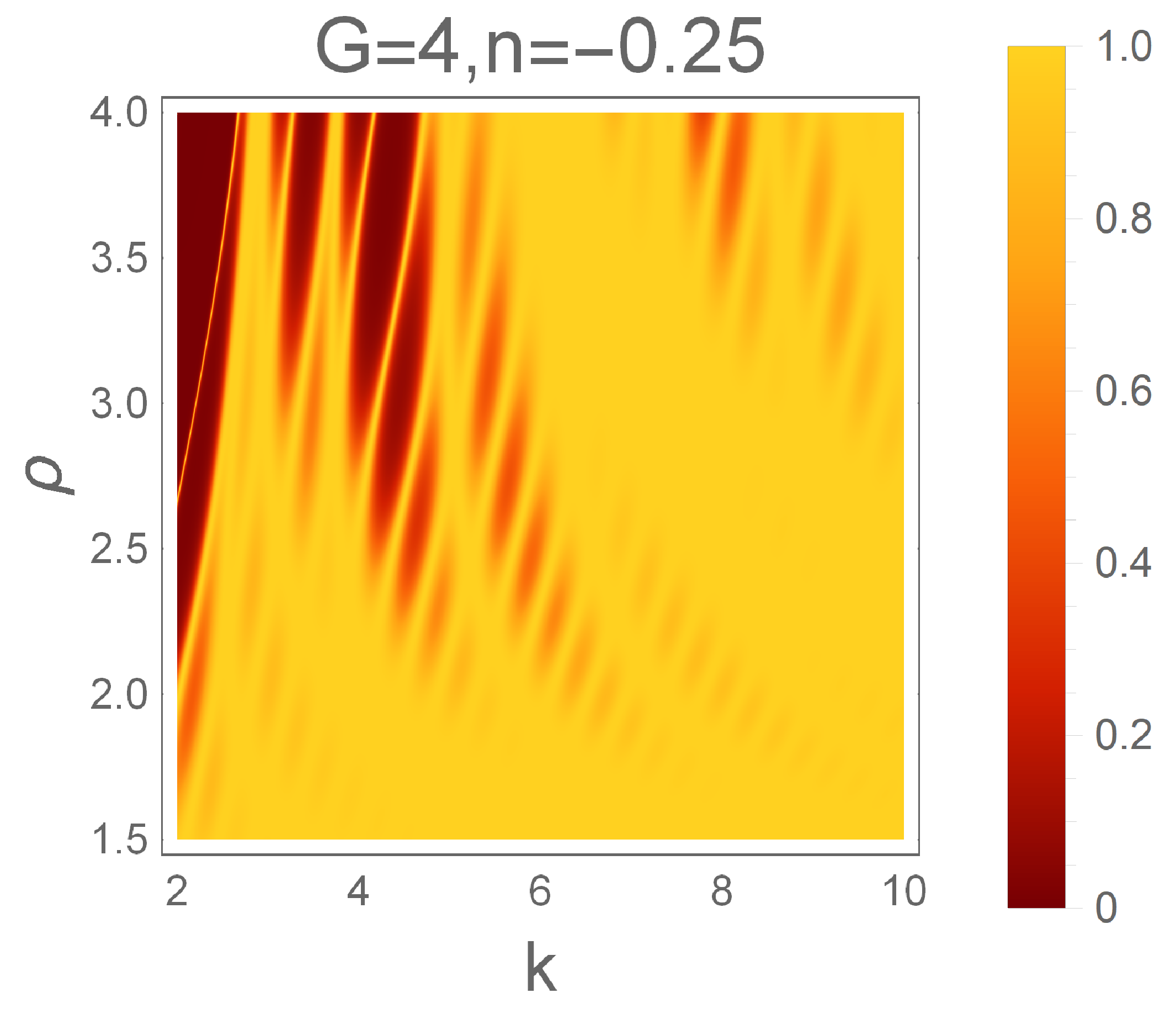} c \\ \includegraphics[scale=0.169]{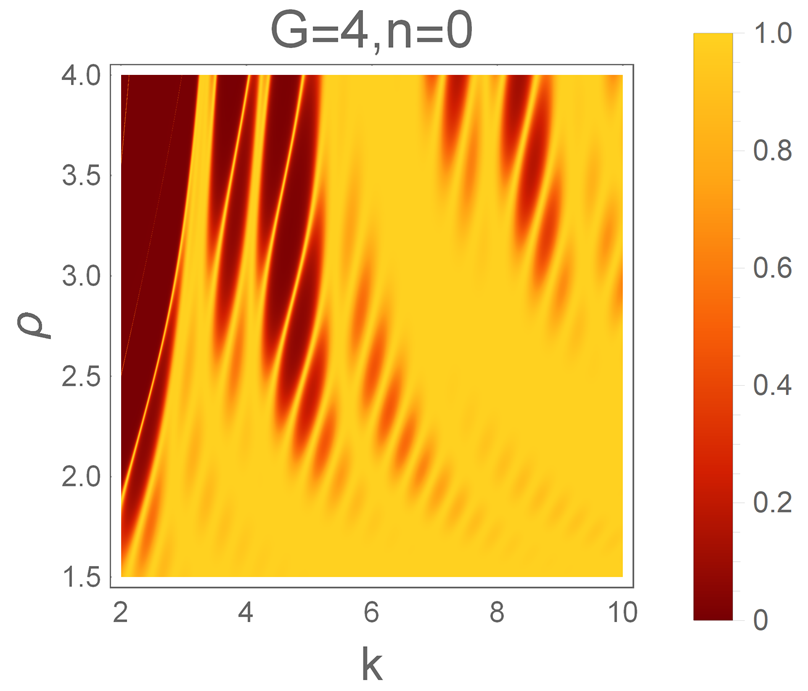} d 
\includegraphics[scale=0.065]{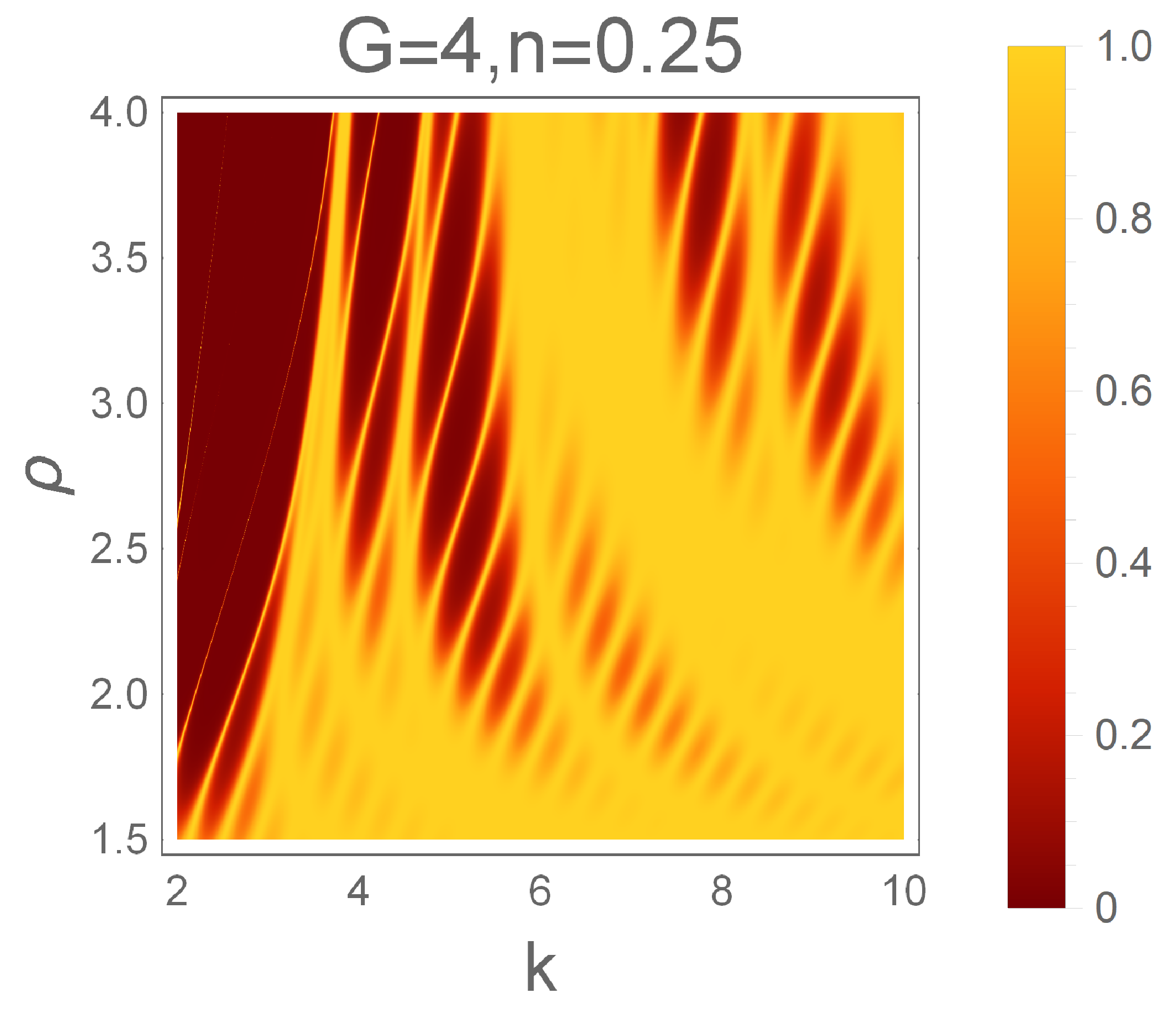}	e 
\includegraphics[scale=0.065]{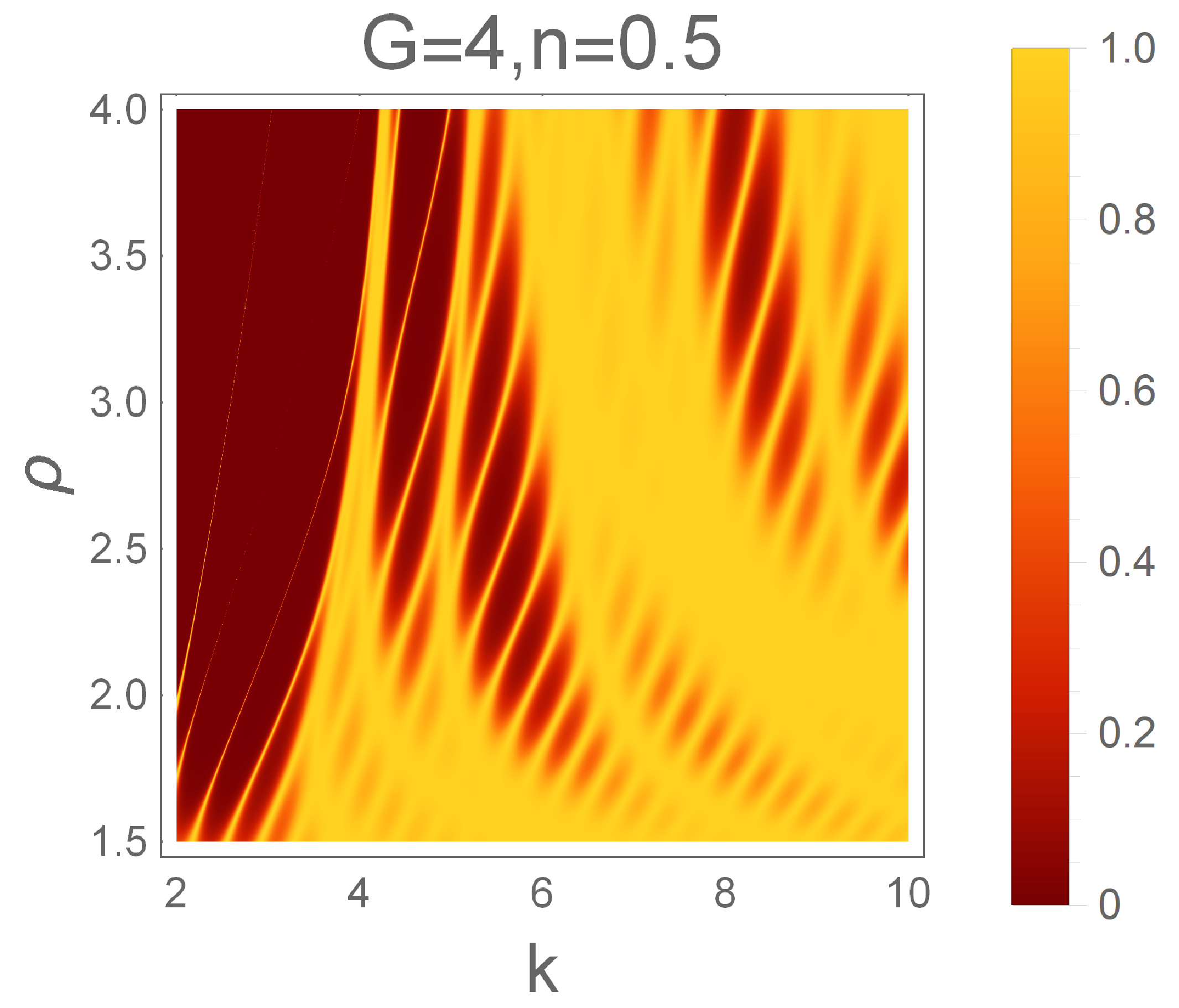} f \\ 
\includegraphics[scale=0.065]{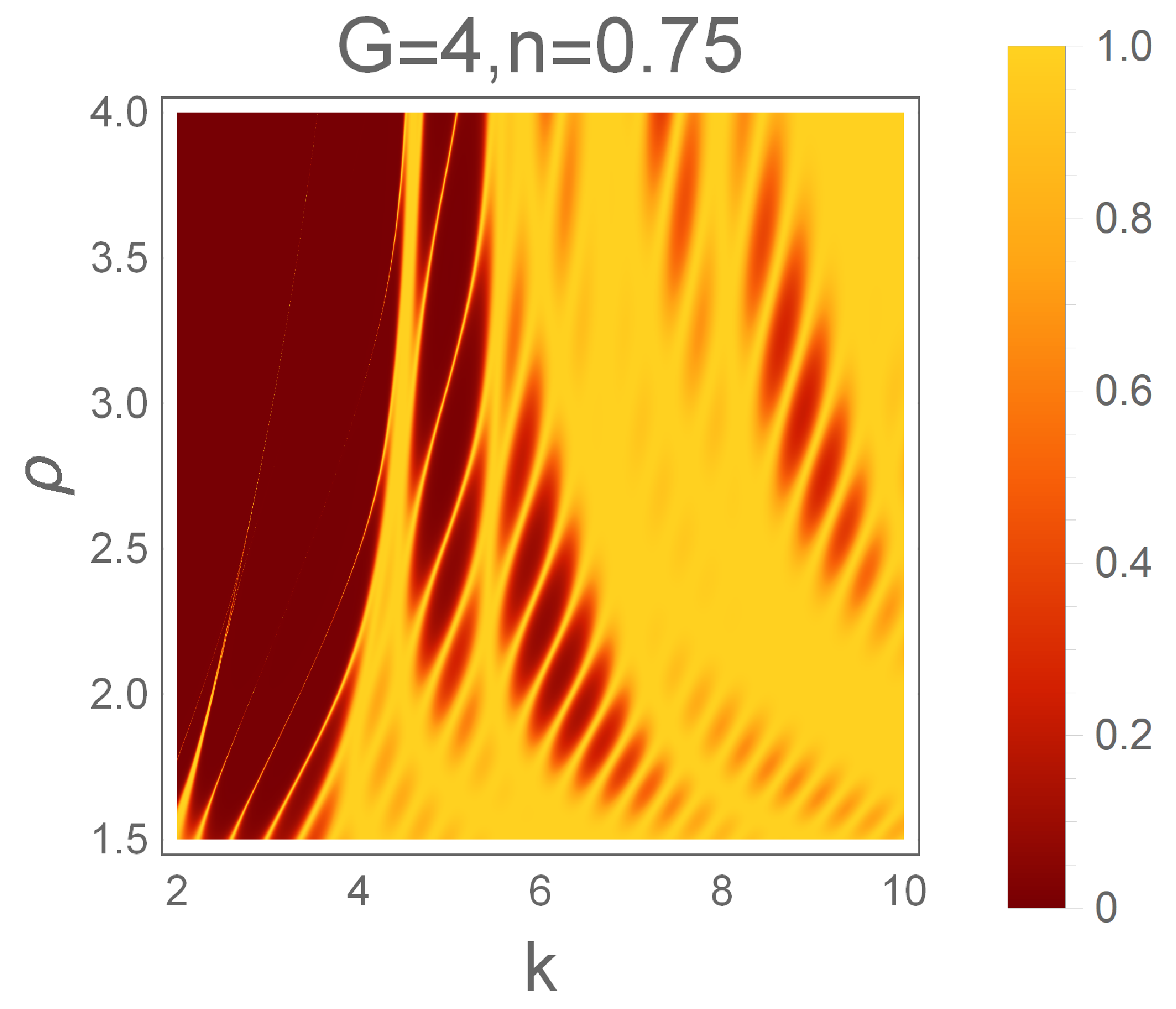} g
\includegraphics[scale=0.065]{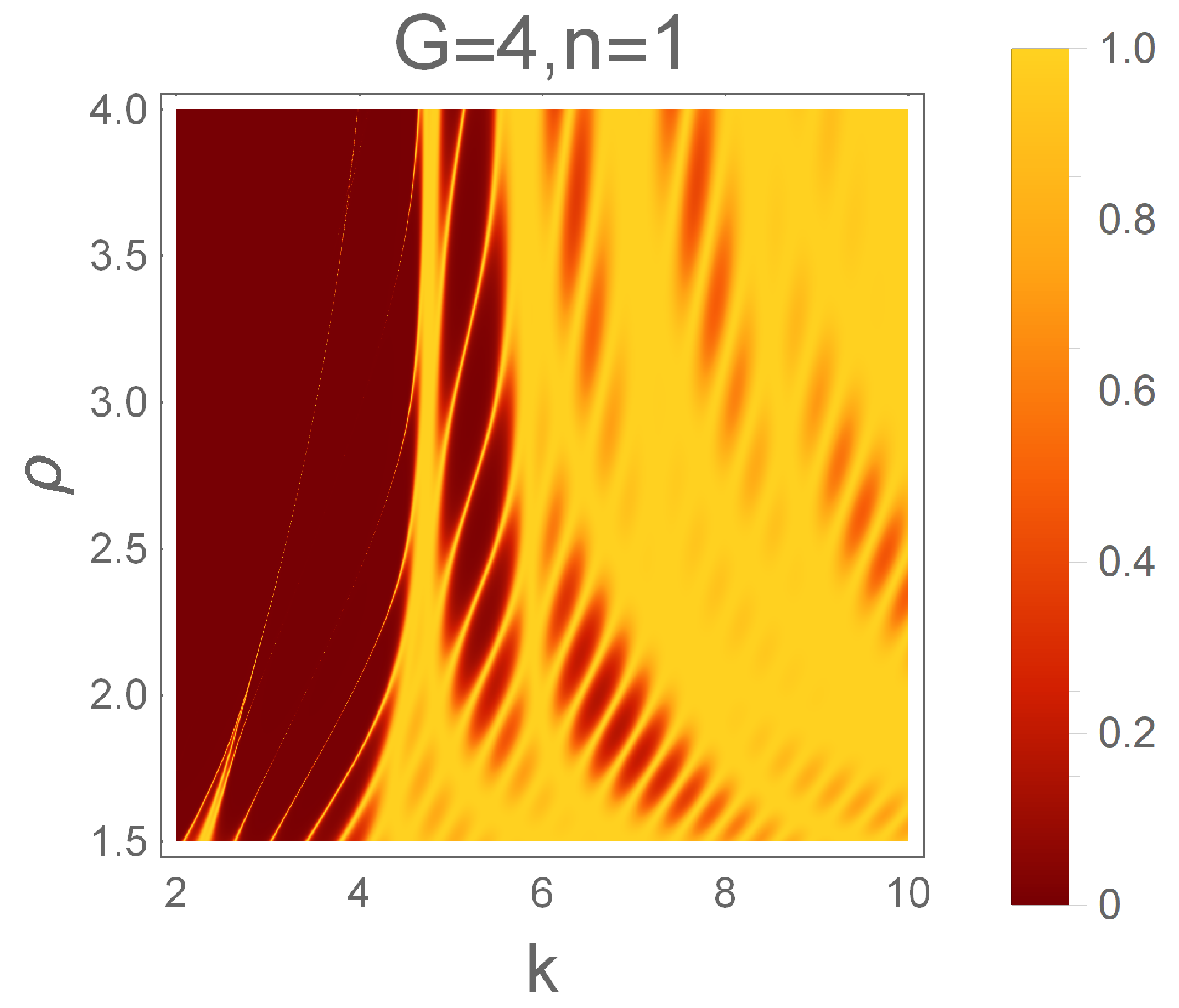} h  
\includegraphics[scale=0.065]{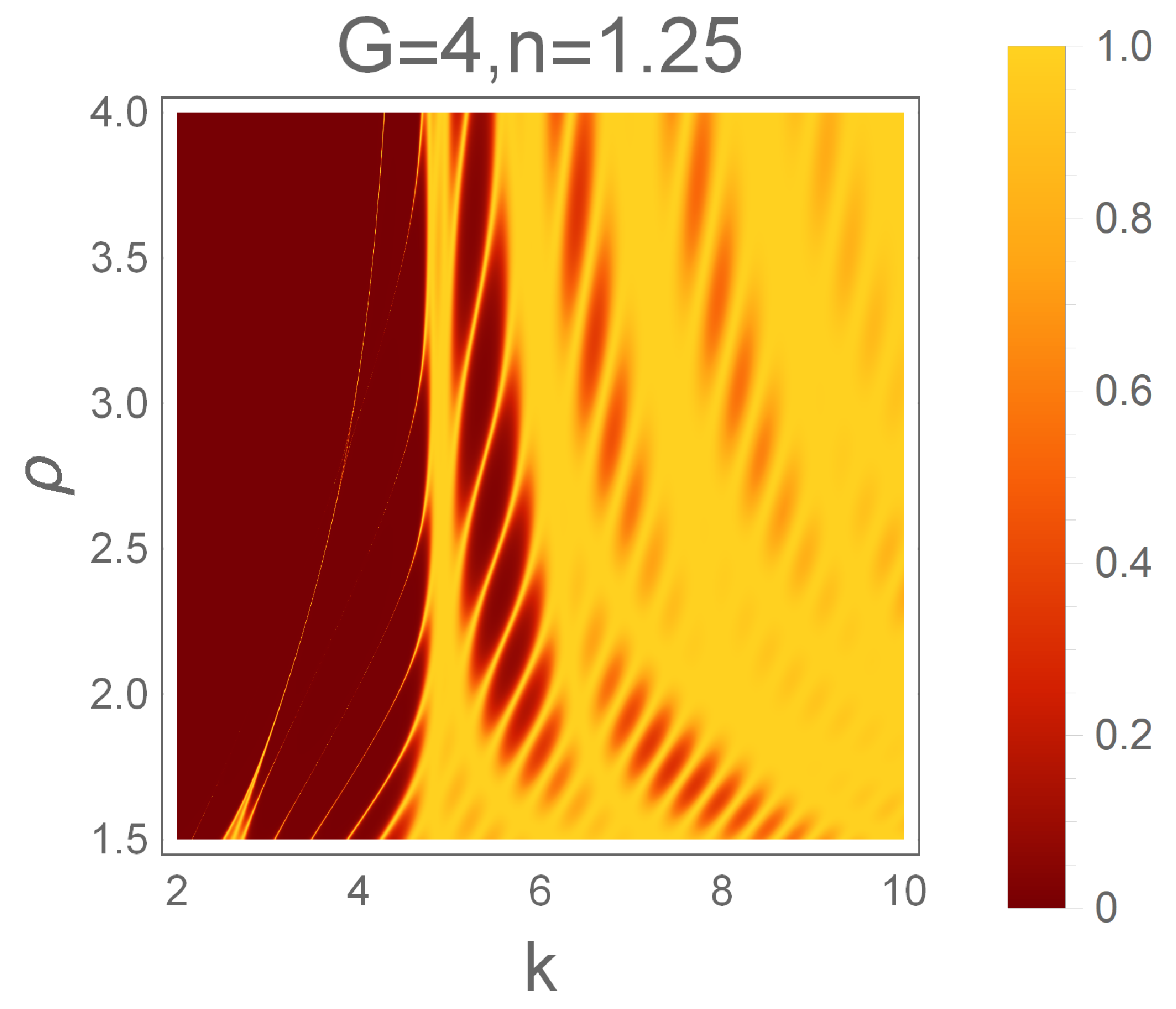} i \\   
\includegraphics[scale=0.065]{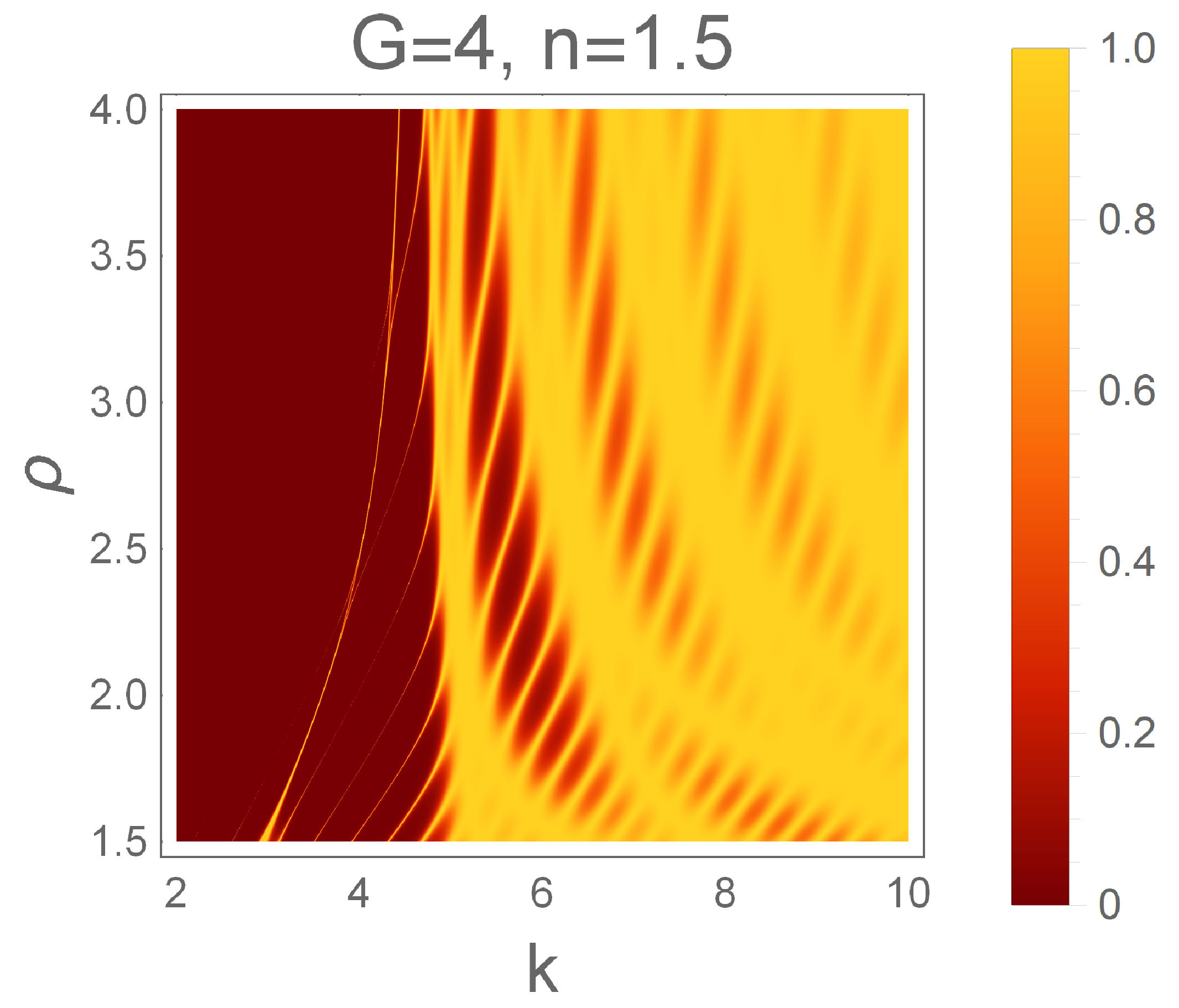} j
\includegraphics[scale=0.065]{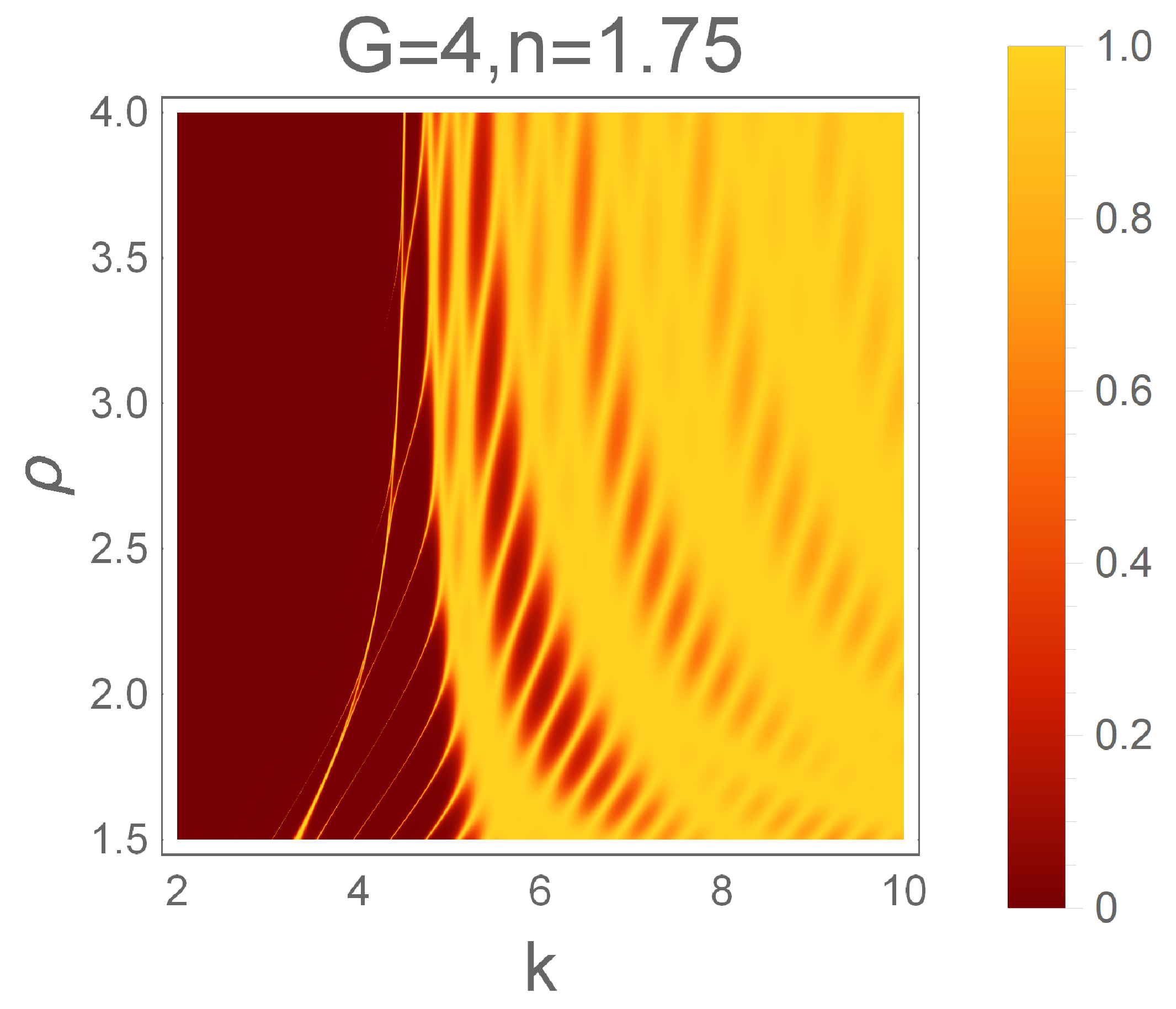} k 
\includegraphics[scale=0.065]{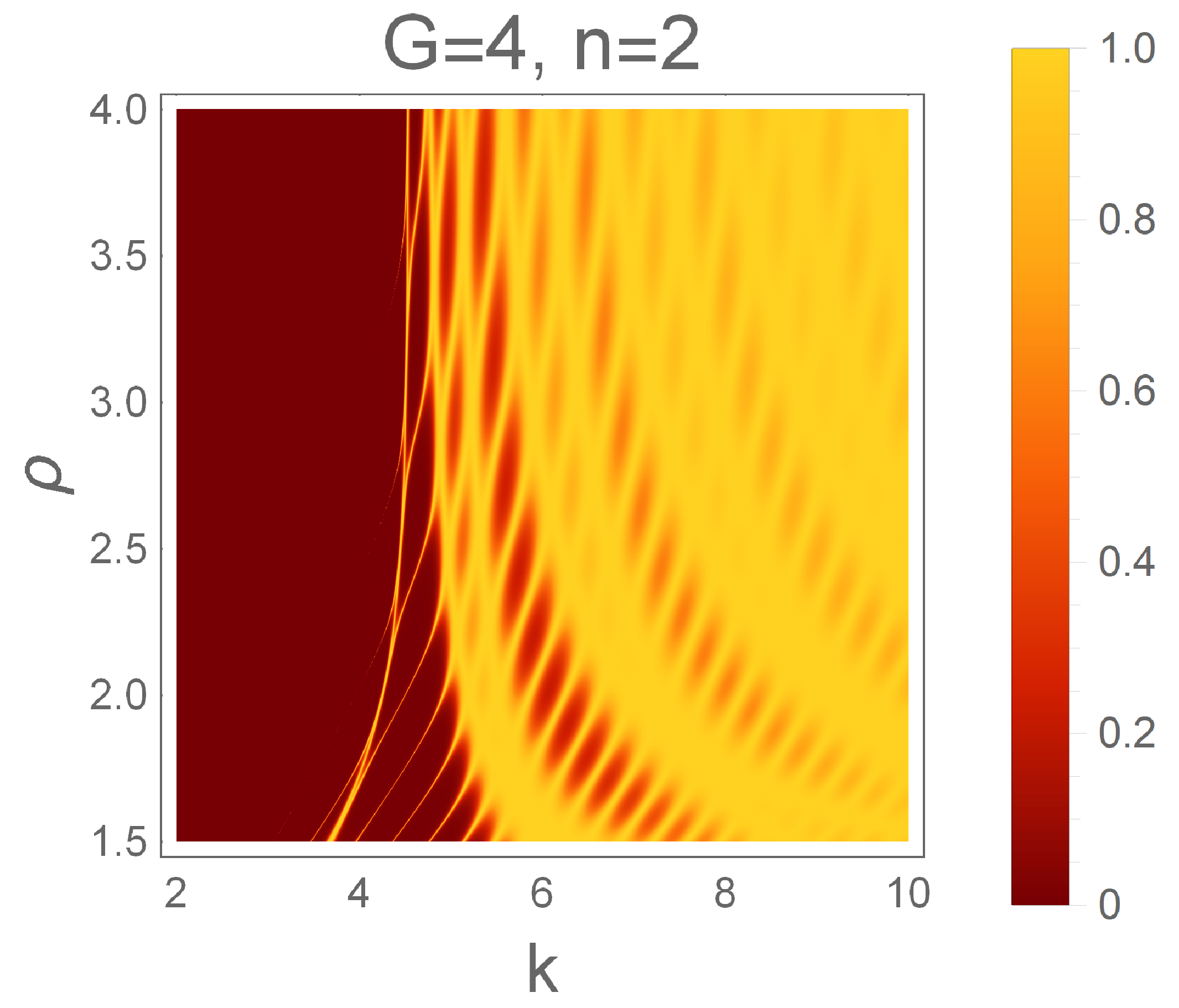} l 
\caption{\it Density plots of tunneling probability $T$ in $\rho-k$ plane for SVC($\rho,n$) potential of different values of $n$ for stage $G=4$. All other potential parameters are the same as of Fig \ref{dp_nk_g3}. Again it is seen that comparatively sharper features of transmission resonance occur for lower $k$ values as $n$ increases.}  
\label{dp_nk_g4}
\end{center}
\end{figure}  
Figs. \ref{dp_nk_g3} and \ref{dp_nk_g4} depicts density plots over $\rho -k$ plane showcasing the behavior of transmission probability $T$ for potential systems characterized by parameters $V=10$ and $L=10$ for stage $G=3$ and $G=4$, respectively. The density plots span a continuous range of $\rho$ from $1.5$ to $4$ and $k$ from $2$ to $10$, with discrete values of $n$ ranging from $-0.75$ to $2$ in increments of $0.25$. The transmission resonances, marked by ultra-thin yellow streaks, represent an abrupt shift of \( T \) from near zero to 1. These resonances are especially discernible between \( k \) values of 2 to 5, increasing in a positive slope as \( \rho \) ascends. As the parameter \( n \) increases, these resonances become so refined that their visual capture becomes a challenge in the presented figures. As such, even partial yellow streaks amidst the predominant maroon regions should be recognized as a continuous resonance that might not be completely visualized due to its sheer fineness.
\par
For both sets of plots, the transmission resonances are prominent for \( n \) values in the positive domain. The intensification of these resonances into ultra-fine streaks with growing \( n \) values is evident. This observation underscores the nuanced interplay between the parameters \( \rho \), \( k \), and \( n \). The parameter $n$, being a new element in this research, plays a pivotal role in defining the resonance behavior of the system. For the \( \rho \) parameter,  as \( \rho \) increases, sudden spikes or resonances in transmission (thin yellow streaks) become prominent over the low transmission regions (red regions). The transition point where these resonances become evident notably shifts with variations in the parameter \( n \). The spacing, initiation point, and complexity of transmission resonances with respect to $k$ vary between $G=3$ and $G=4$. 
\par
\begin{figure}[h! tbp]
    \centering
\includegraphics[scale=0.147]{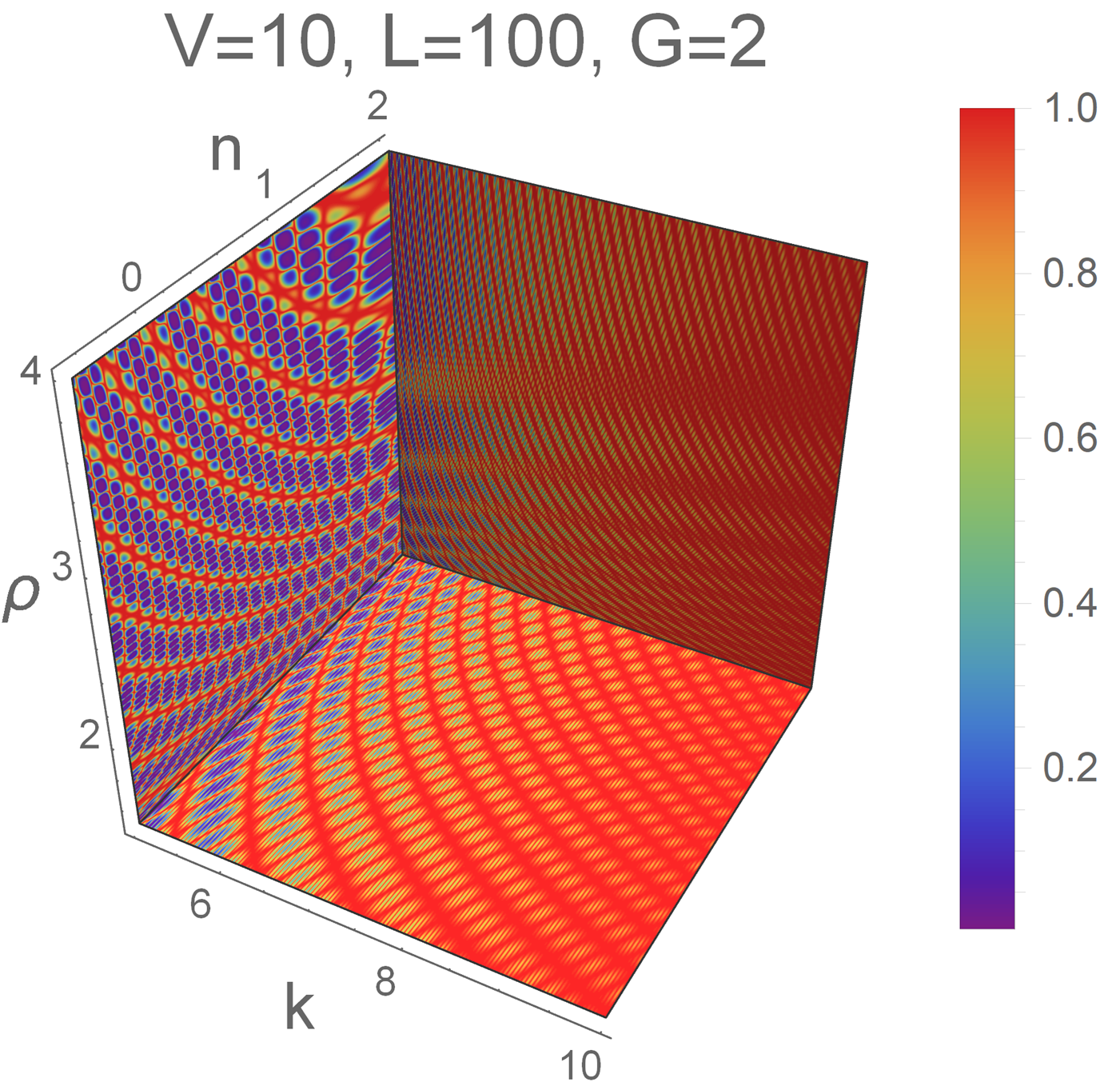} a
\includegraphics[scale=0.147]{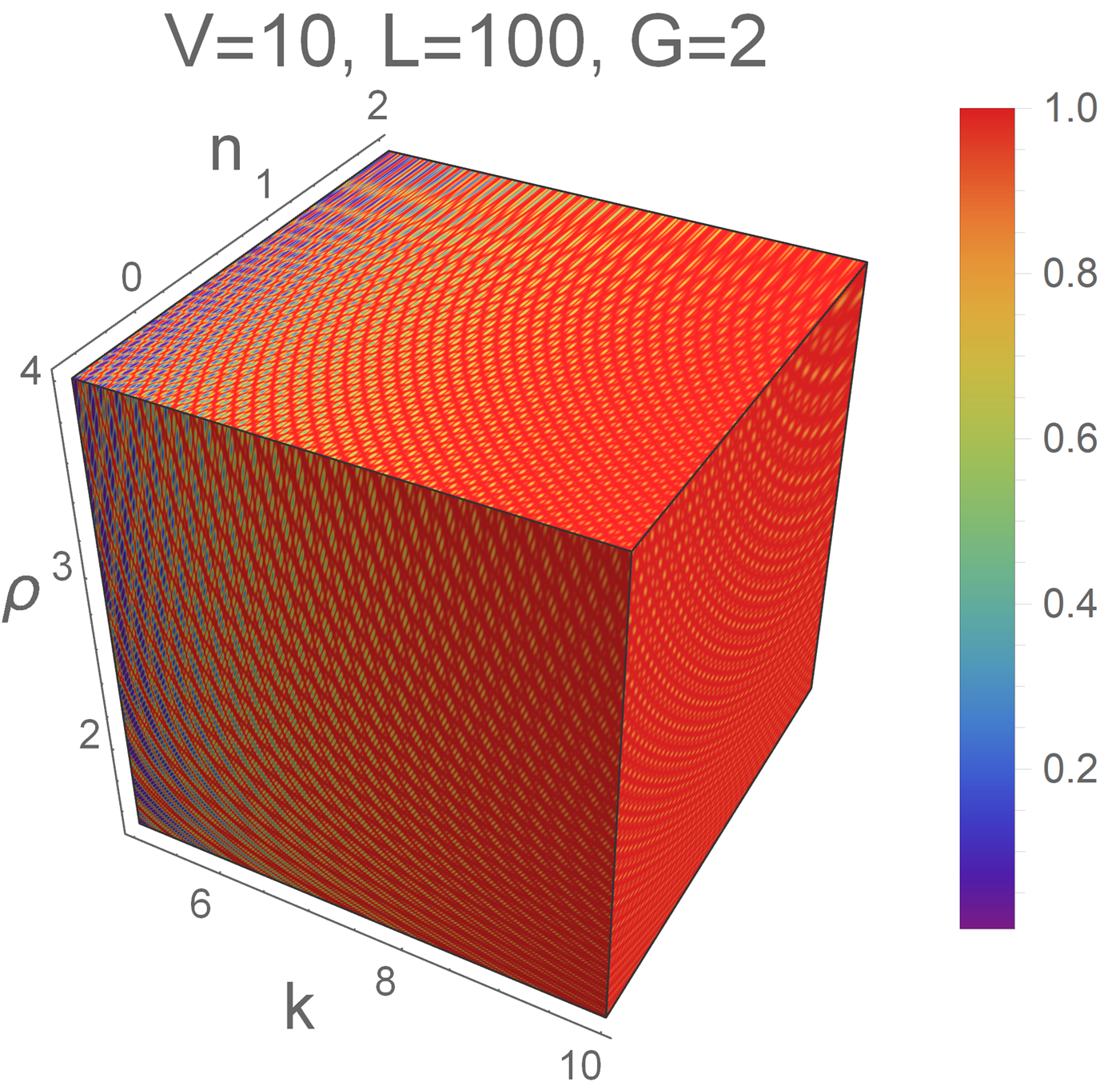} b\\

\includegraphics[scale=0.212]{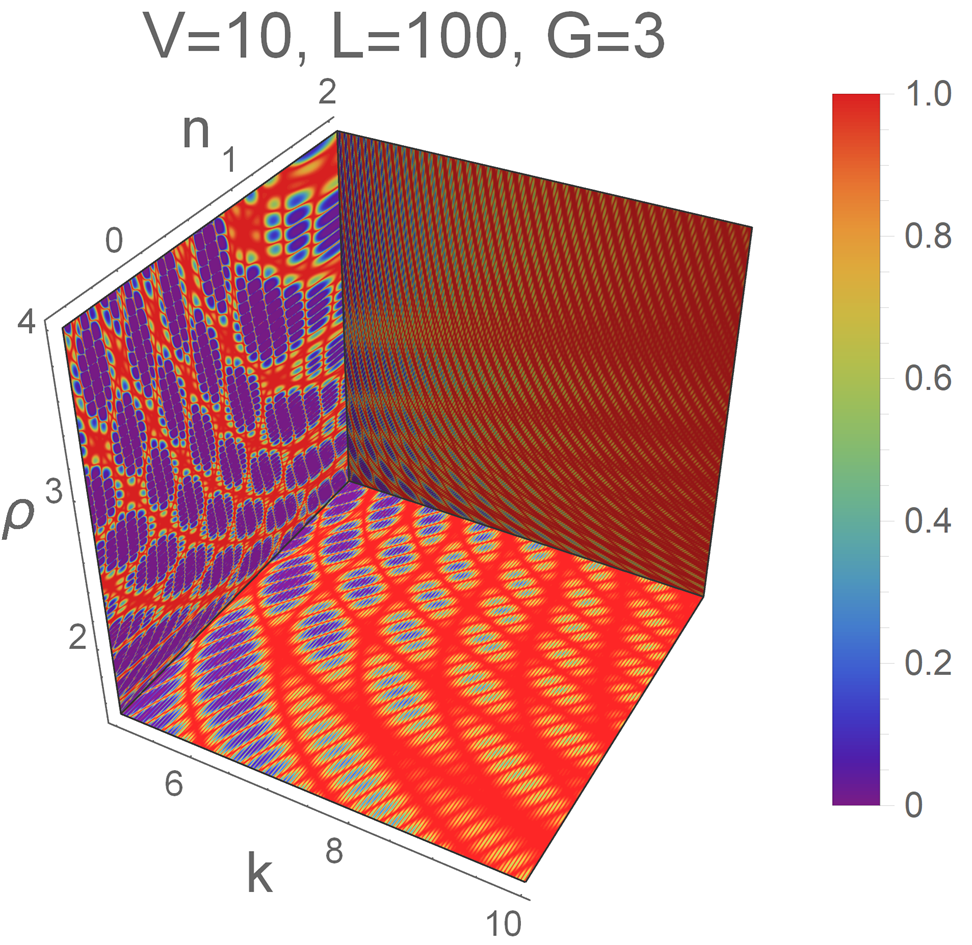} c
\includegraphics[scale=0.212]{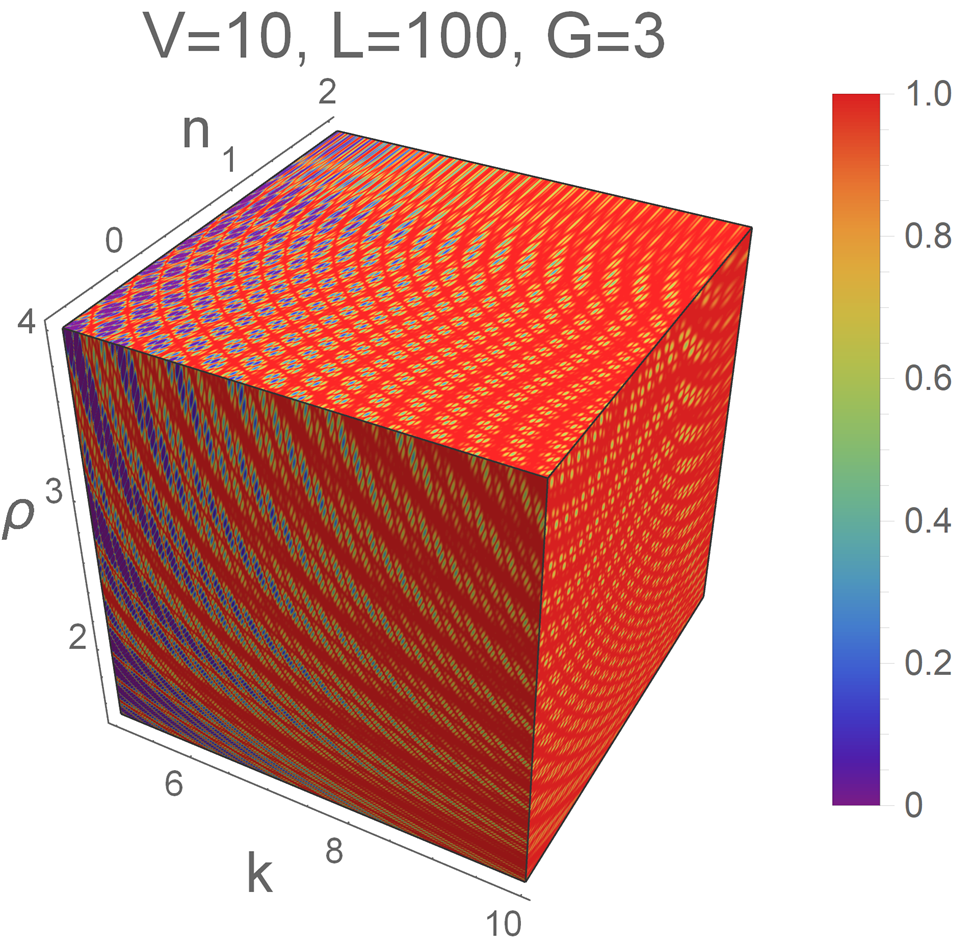} d\\
\includegraphics[scale=0.145]{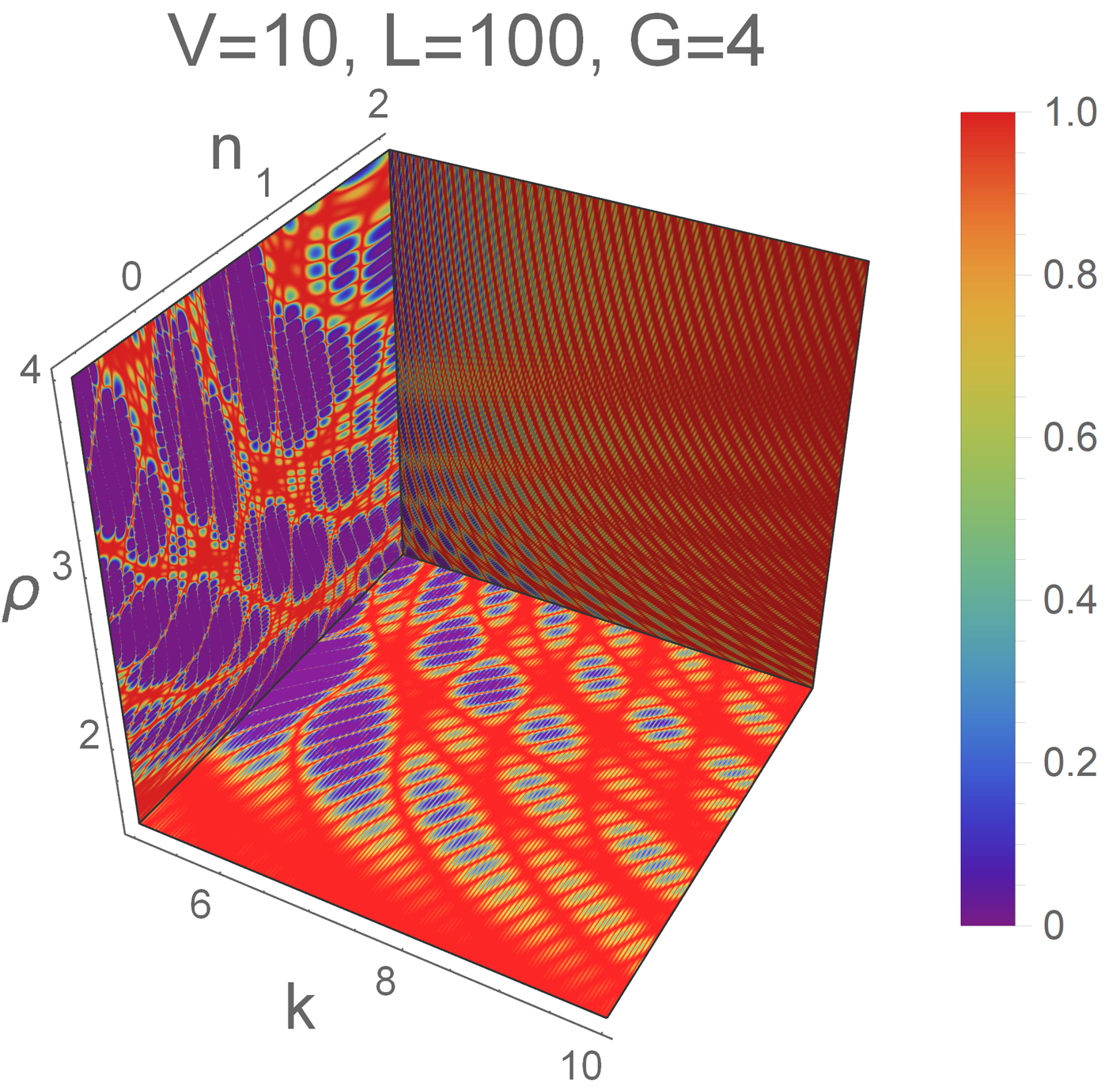} e
\includegraphics[scale=0.145]{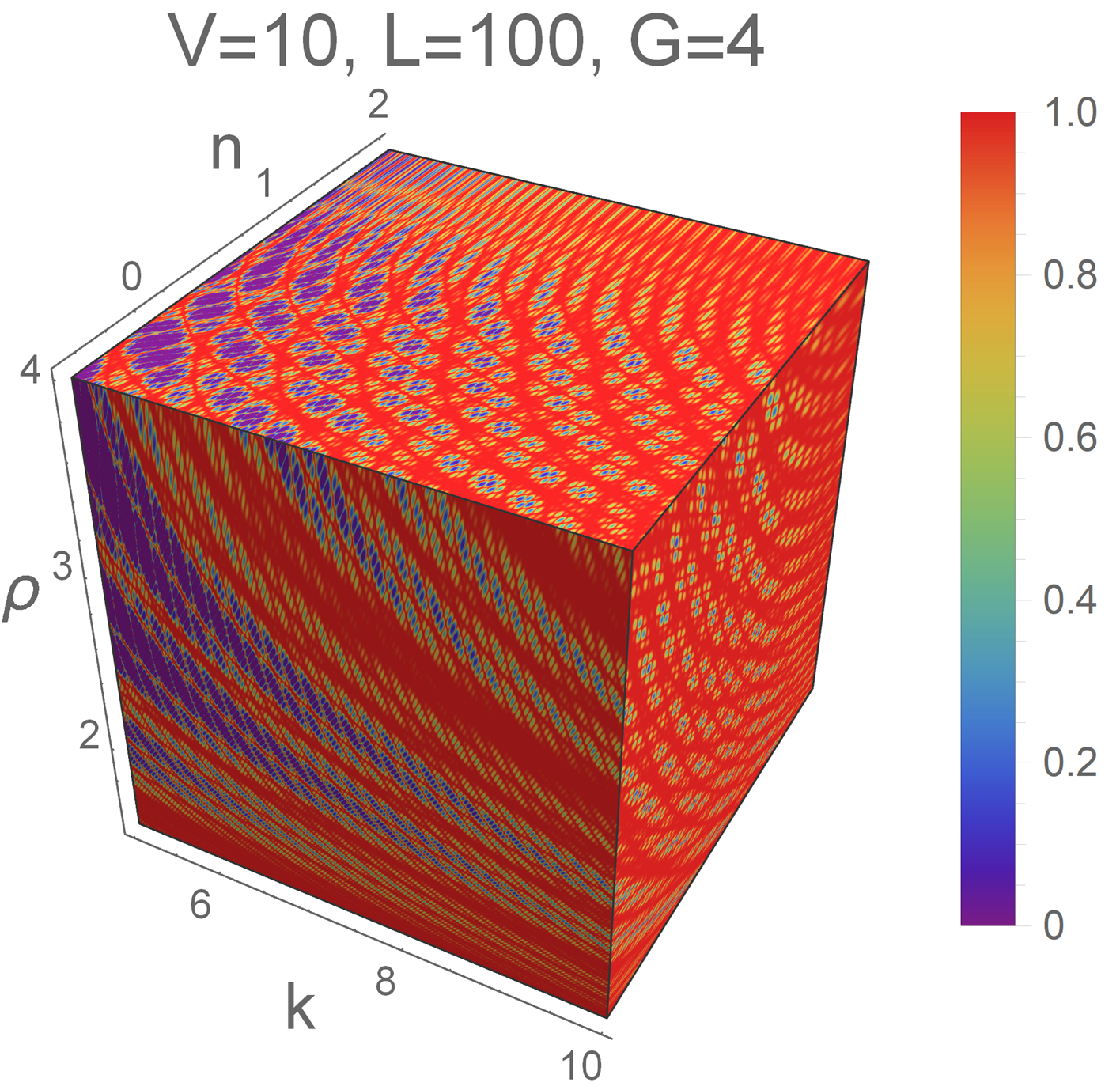} f
\caption{ \it Plot showing density plots of $T$ for different faces of cuboid spanned by $\rho$, $k$ and $n$. Potential parameters are shown in the figures.}
\label{3dDp_G_plane}
\end{figure}

To present the more intricate and rich features of tunneling from SVC$(\rho,n)$ potential systems, Fig. \ref{3dDp_G_plane} presents density plots showcasing transmission probabilities for various stages ranging from $G=2$ to $G=4$. The density plots of $T$ are shown over the front and back plane of the cuboid spanned by $\rho, k$ and $n$. At a glance, one can discern the intricate patterns that progressively evolve with increasing $G$ values. Very rich patterns emerge from sparser, ordered configurations to densely clustered regions of high and low transmission as $G$ increases, signifying the profound influence of more divisions in the potential system. Extremely fine streaks of red lines show very sharp transmission resonances from this potential. Over $\rho-n$ plane, in the initial stages, $G=2,3$ the transmission regions are distinguished by their more isolated and sporadic configurations. These manifest as discrete islands representing higher or lower transmission probabilities. However, as the division process progresses to $G=4$, there is a further increase in the sharpness of transmission resonance and the intricacy of the patterns. The transmission regions transform into more tightly knit, overlapping configurations, suggesting a greater influence of the parameter $n$. Similarly, the structures are more pronounced over the $k-n$ plane with increasing $G$ which indicates more extreme behavior of $T$ with the increase in stage $G$ of the system. This indicates that SVC$(\rho, n)$ potential is more rich system as compared to the general SVC$(\rho)$ potential system.  Overall the figure indicates that the transmission dynamics gains layers of unpredictability with an increased number of divisions of the potential, making a case for further computational or analytical scrutiny to grasp the underlying mechanisms and their broader implications.
\par
\begin{figure}[H]
\centering
\includegraphics[scale=0.47]{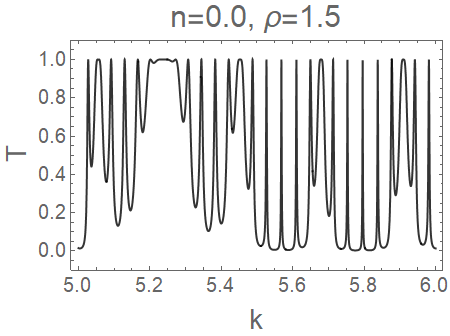} a
\includegraphics[scale=0.47]{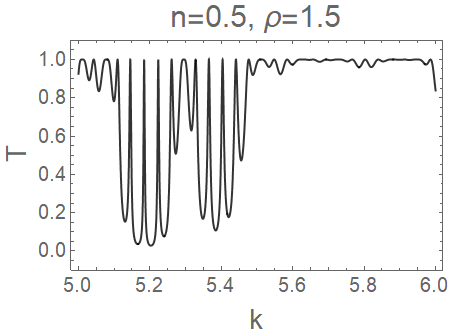} b \\
\includegraphics[scale=0.7]{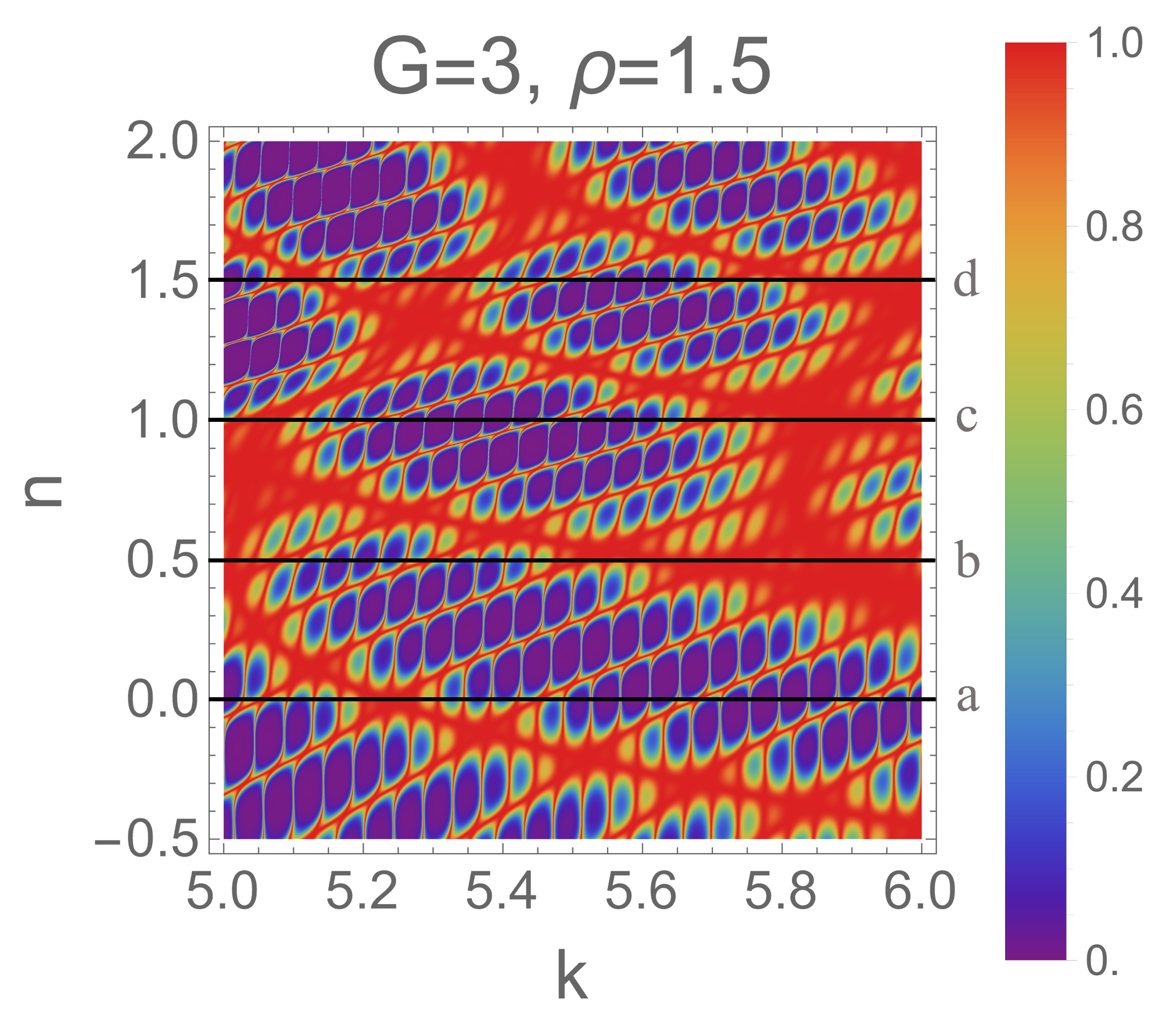} \\
\includegraphics[scale=0.47]{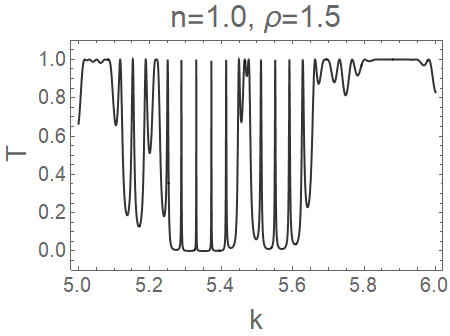} c
\includegraphics[scale=0.47]{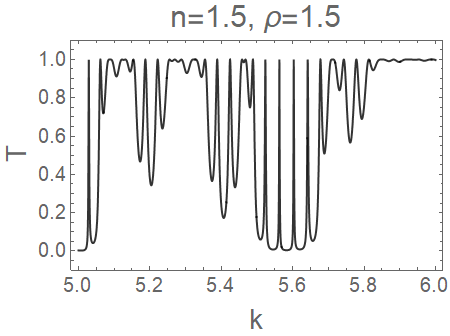} d
\caption{ \it Density plot showing the distribution of transmission probability T within the $n-k$ plane. The depicted density plane is a segment of Fig. \ref{3dDp_G_plane}c for a small range of k, using identical system parameters. The surrounding four 2D plots a, b, c and d show the variation of transmission probability T with k, ranging from $5$ to $6$, for $n=0$, $0.5$, $1.0$ and $1.5$ respectively. Black lines on the density plots indexed as a, b, c and d correspond to surrounded 2D plots respectively.}
    \label{DP_BP_2d_01}
\end{figure}
\begin{figure}[H]
\centering
\includegraphics[scale=0.47]{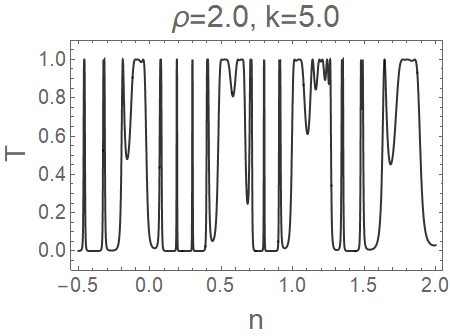} a
\includegraphics[scale=0.47]{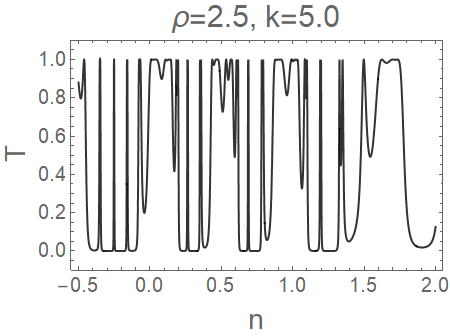} b \\
\includegraphics[scale=0.67]{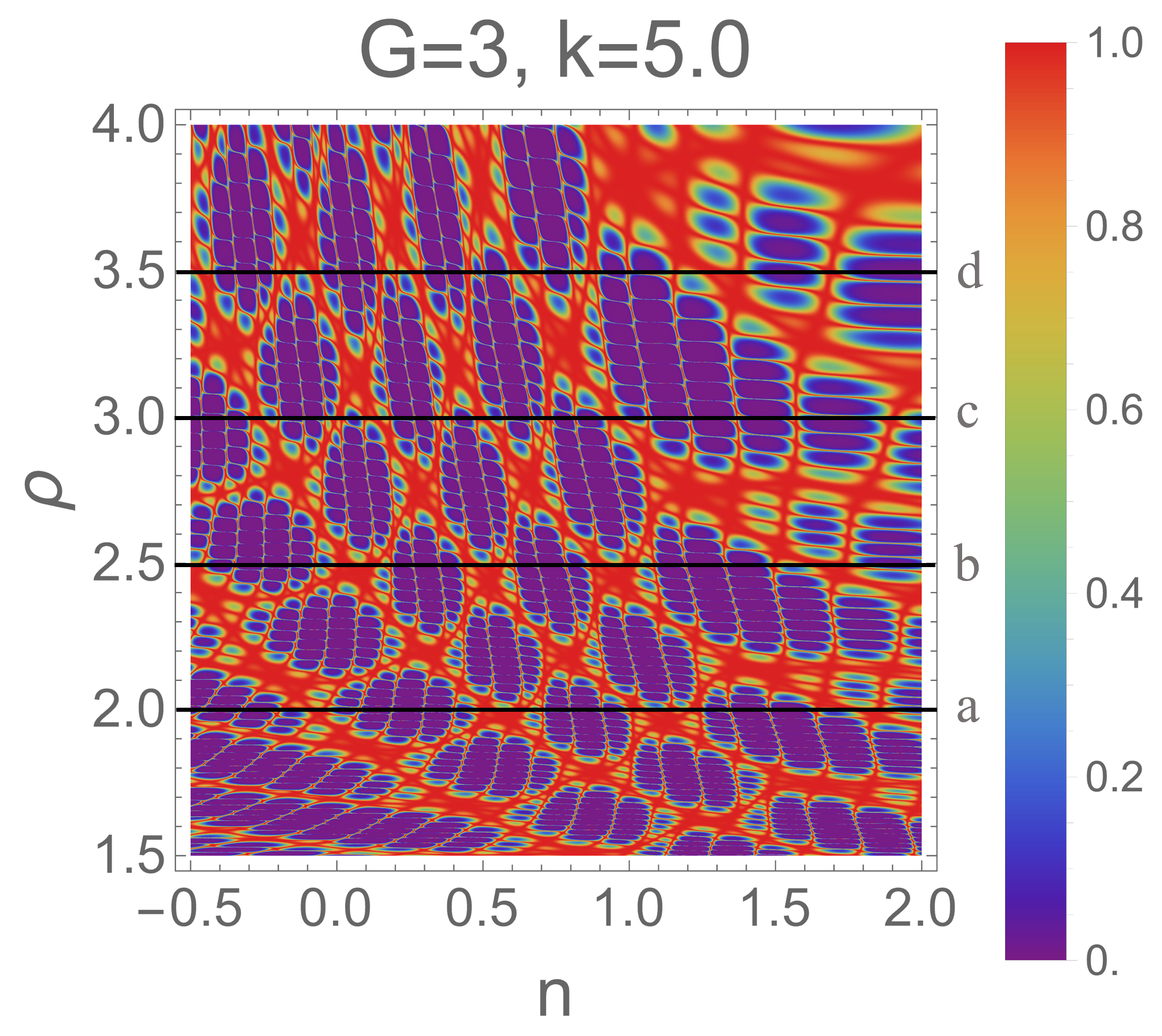} \\
\includegraphics[scale=0.47]{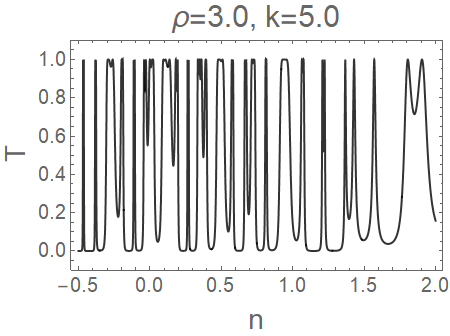} c
\includegraphics[scale=0.47]{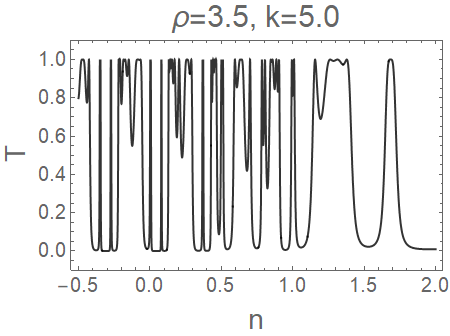} d
\caption{\it Density plot showing the distribution of transmission probability T within the $\rho-n$ plane. The depicted density plane is a segment of Fig. \ref{3dDp_G_plane}c with same identical system parameters. The surrounding four 2D plots a, b, c and d show the variation of transmission probability T with n, ranging from $-0.5$ to $2.0$, for $\rho=2.0$, $2.5$, $3.0$ and $3.5$ respectively. Black lines on the density plots indexed as a, b, c and d correspond to surrounded 2D plots respectively.}
\label{DP_BP_2d_02}
\end{figure}
In Fig. \ref{DP_BP_2d_01}, we show the density plot of $T$ against wave number $k$ and parameter $n$. For the potential system illustrated, the characteristics correspond to stage $G=3$, $\rho=1.5, V=10$ and $L=100$. Our color gradient, which transitions from violet to red, provides a visualization of tunneling probabilities ranging from 0 (total reflection) to 1 (full transmission).  Flanking the central density plot, four 2D plots lay out the behavior of \( T \) as a function of \( k \) for distinct \( n \) values: 0.0, 0.5, 1.0, and 1.5. Black guide lines are drawn over the density plot corresponding to these $n$ values. These plots capture the rich transmission structure with varying $n$. More analytical studies would be needed to characterize these transmission features. At \( n=0.0 \) and \( n=1.0 \), we notice the prevalence of very sharp transmission resonances, hinting at heightened tunneling sensitivity for this system. As we shift to \( n=0.5 \) and \( n=1.5 \), the transmission resonances are less in numbers. This observation with $n$ underscores the intricate dependency on the \( n \) parameter. In the central density plot, as \( n \) increases, there's a noticeable shift in the tunneling probability pattern. The valleys, indicative of lower tunneling probabilities, vary in their width and depth. As \( n \) progresses, the width of these valleys undergoes changes, and there's a discernible difference in the depth of these valleys, suggesting regions of stronger and weaker suppression of tunneling. The frequency of these valleys also greatly modulated with \( n \), indicating varying oscillations in tunneling probability across the range.
\par
The purpose of Fig. \ref{DP_BP_2d_02} is to demonstrate the tunneling features for a fixed wave number $k$ ($k=5$ here) over varying $\rho-n$ planes. The figure shows the density plot of $T$ for stage $G=3$, $V=10, L=100$. It is seen from the figure that over the given $\rho-n$ plane, the tunneling probability clearly varies from near 0 to 1 ($T=0$ is never possible for the Hermitian system \cite{mostafazadeh2018_scattering}). This indicates even for a fix span $L$ ($L=100$ at present) for the potential system, we may achieve any desired value of $T$ by tuning the parameter $\rho$ and $n$. This shows the versatile nature of this system. Again, the presence of thin streaks of red lines in this figure highlights specific regions where tunneling probabilities peak, showing very sharp transmission resonance even at a fixed wave number \( k \). As we traverse along the \( n \) axis, the resonance patterns undergo discernible transformations, emphasizing the sensitivity of tunneling phenomena to this parameter. Rapid changes in $T$ is observed for smaller $n$ values in the figure. For comparatively smaller $\rho$ and $n$ values (lower left in the figure), the extreme variation in $T$ is more rapid as compared to larger $\rho$ and $n$ values (upper right in the figure). The adjacent plots, corresponding to specific values of \( \rho = 2\), 2.5, 3 and 3.5, offer a magnified insight into the behavior of the tunneling probability. These plots, each demarcated by black lines on the central density plot, capture the intricate resonance structures as a function of \( n \). As \( \rho \) increases, there's a perceptible evolution in the resonance patterns. While the fundamental characteristics of tunneling resonances remain, the spacing, amplitude, and shape exhibit variability across different \( \rho \) values. This underlines the influence of \( \rho \) in modulating the tunneling phenomena at fixed wave-number $k$, showing again the richness of the system's response to parameter variations.
\section{Saturation of tunneling profile with parameter \texorpdfstring{$n$}{n}}
\label{saturationn}
\begin{figure}[H]
\centering
\includegraphics[scale=0.39]{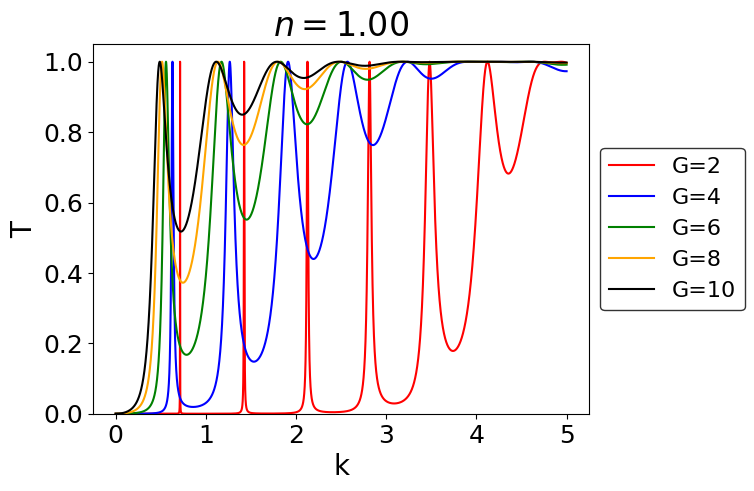} a
\includegraphics[scale=0.39]{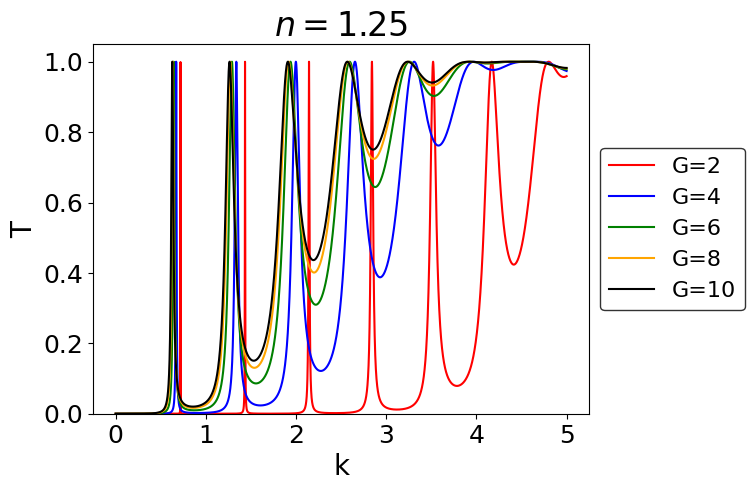} b \\
\includegraphics[scale=0.39]{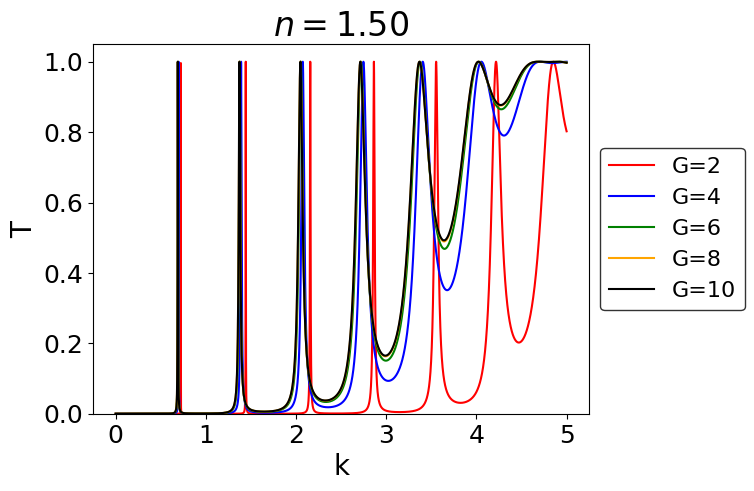} c
\includegraphics[scale=0.39]{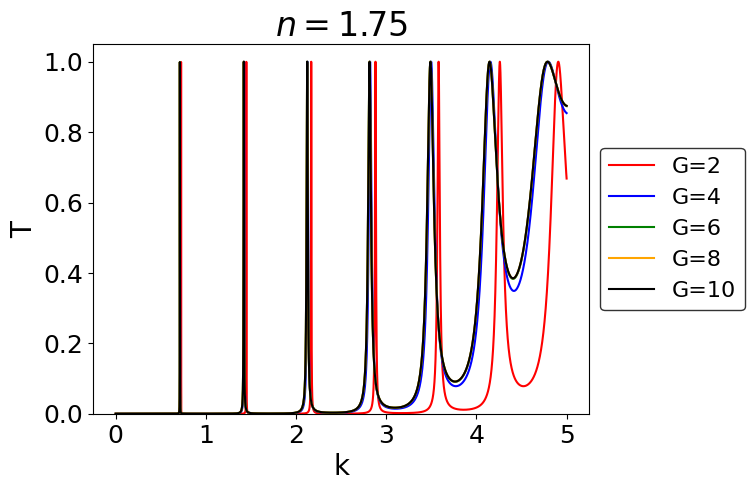} d \\
\includegraphics[scale=0.39]{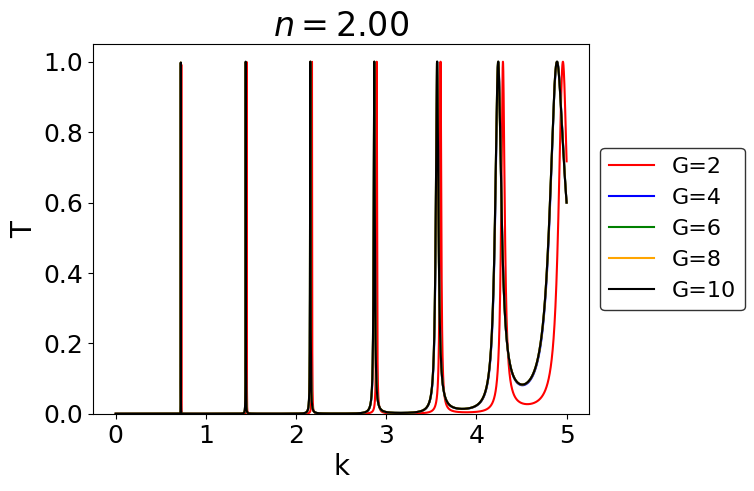} e
\includegraphics[scale=0.39]{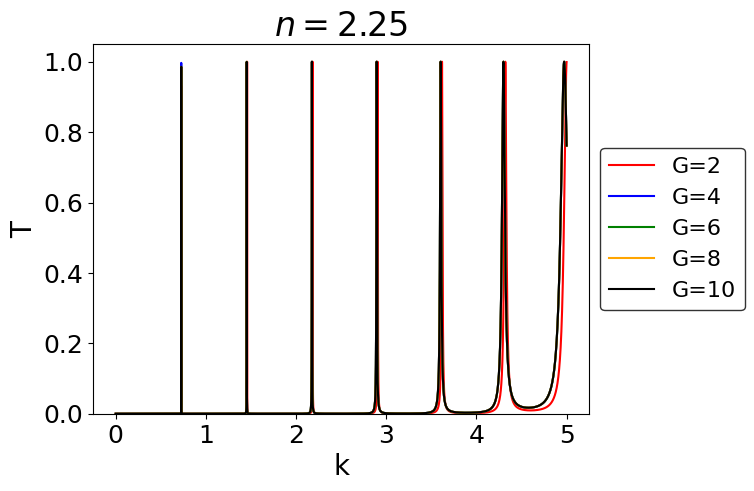} f \\
\includegraphics[scale=0.39]{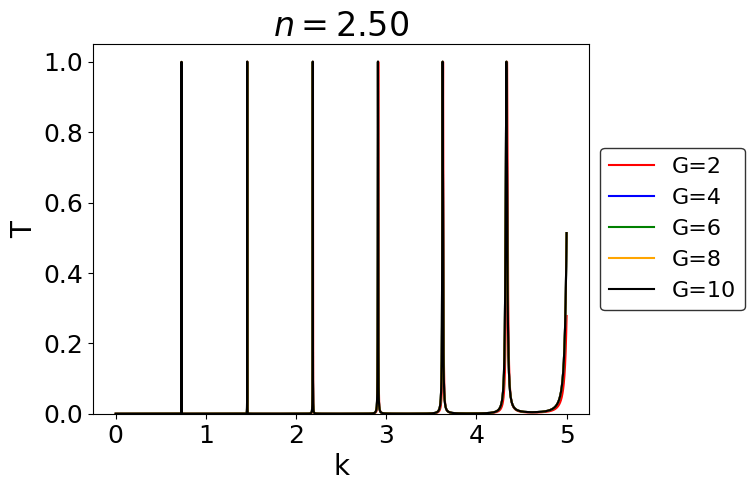} g
\includegraphics[scale=0.39]{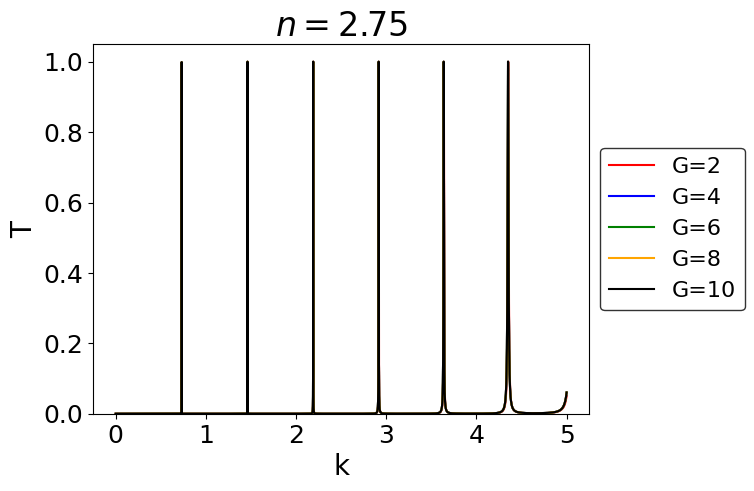} h 
\caption{ \it {Plots showing the saturation of transmission profile as the parameter n increases for a fixed stage}. The potential parameters used are $L =$ $5.0$, $V =$ $25.0$, and $\rho =$ $1.25$ for various values of n, as depicted in the accompanying figure. Notably, the results reveal that as the value of n increases, saturation occurs at relatively smaller values of G.}
\label{sat_G}
\end{figure}
\begin{figure}[H]
\centering
    \includegraphics[scale=0.48]{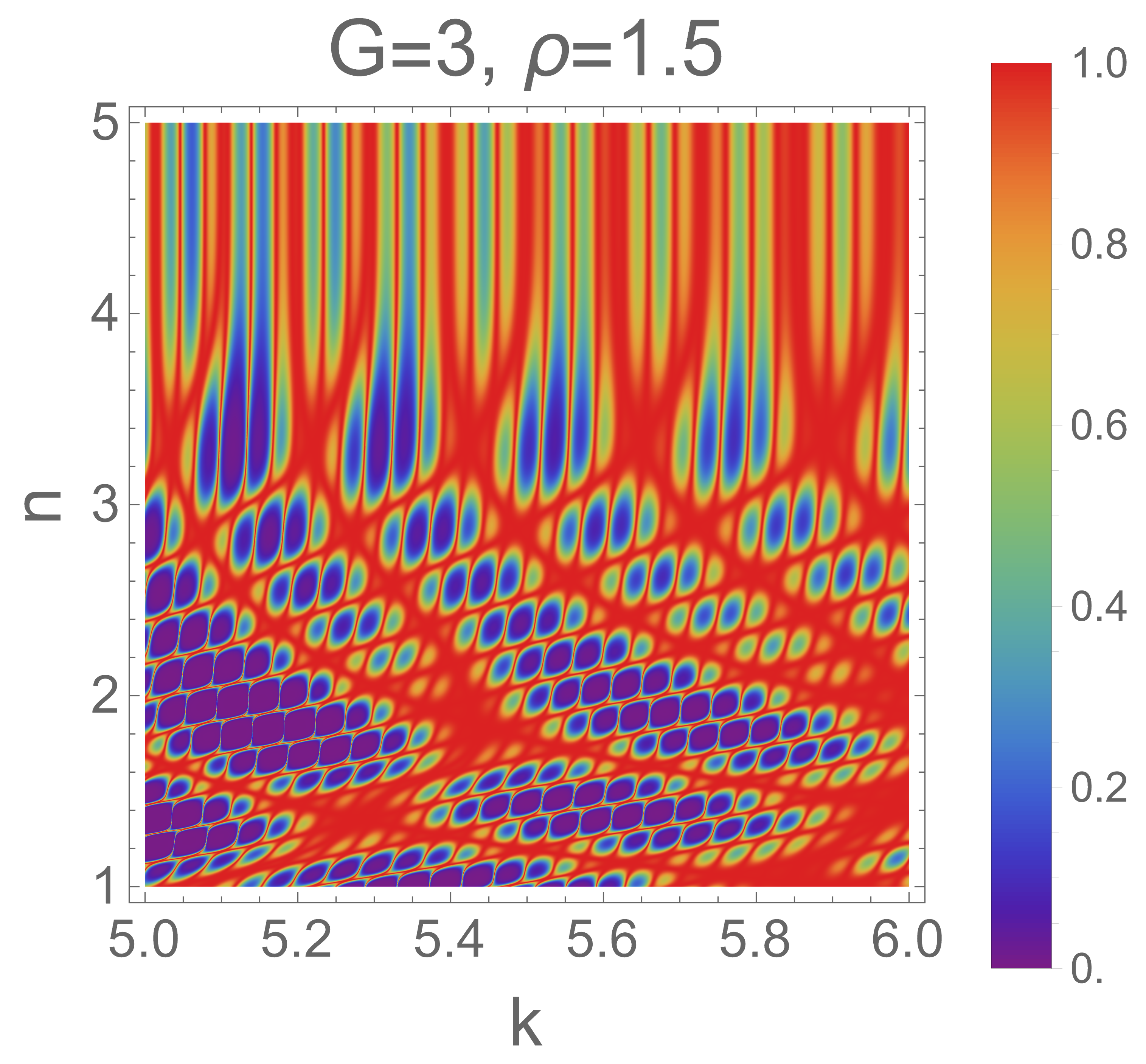} a
    \includegraphics[scale=0.49]{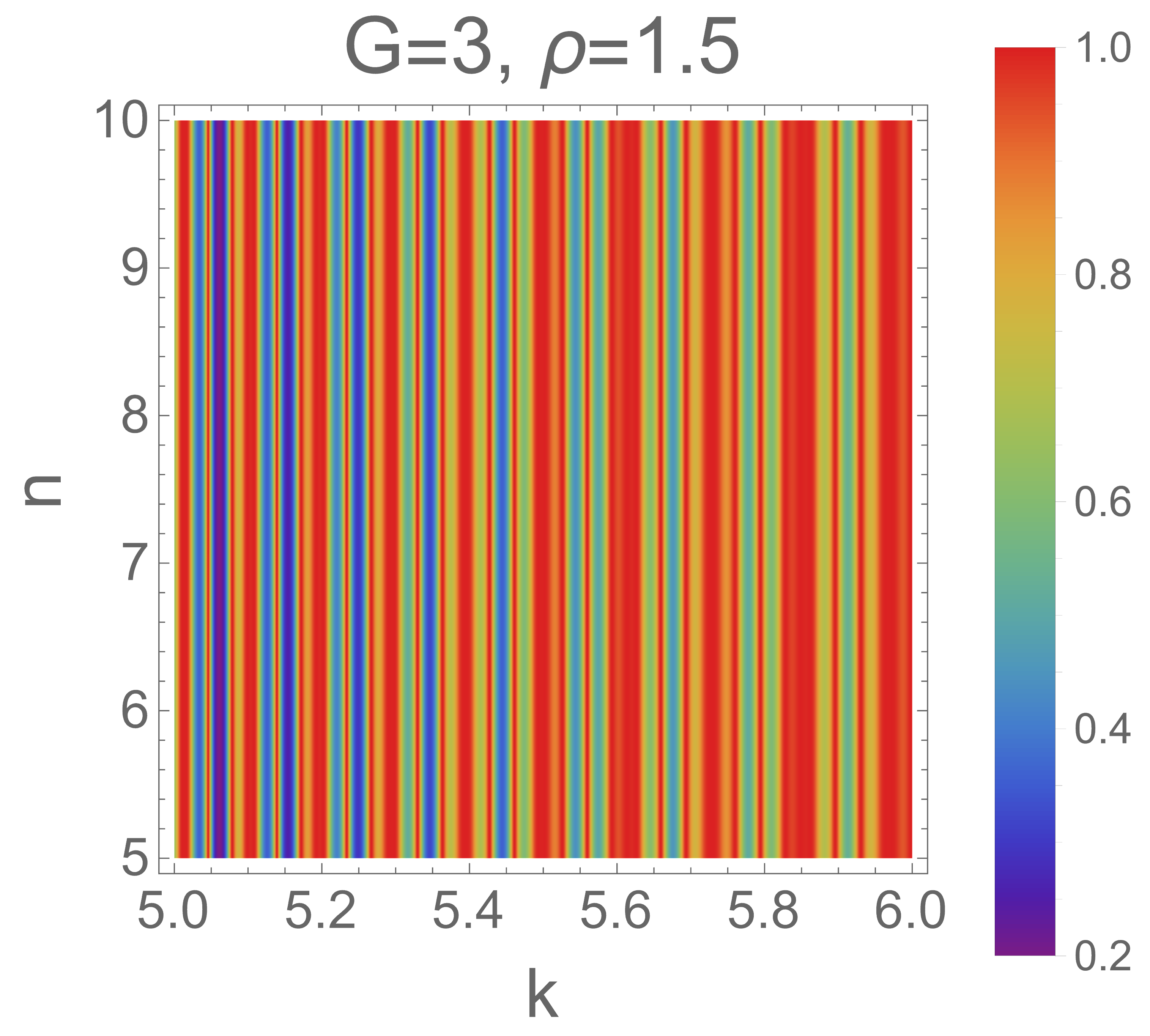} b
    \caption{\it Density plots of the transmission coefficient \( T \) as a function of the wave vector \( k \) and the parameter \( n \) for SVC$(\rho,n)$ potential with parameters \( G = 3 \) and \( \rho = 1.5 \), with a potential height \( V = 10 \) and width \( L = 100 \). (a) The range of \( n \) is from 1 to 5, showcasing the initial variation in \( T \) and the emergence of saturation. Red indicates full transmission (\( T = 1 \)), transitioning to violet for no transmission (\( T = 0 \)). (b) The extended range of \( n \) up to 10, where the pattern of \( T \) exhibits clear saturation, indicating minimal changes in transmission with increasing \( n \). }
    \label{sat_n}
\end{figure}
As $n$ increases, while keeping $G$ fixed, the fraction of the segment removed $\left(\frac{1}{\rho^{G^n}}\right)$ becomes progressively smaller. This is due to the exponent $G^n$ growing larger with an increase in $n$. The effect of removing a smaller and smaller portion of the potential segment means that the overall structure of the potential changes less significantly for an increase in $n$ when $n$ is large. In this scenario, the transmission coefficient $T$, is expected to show saturation as $n$ becomes large. This saturation implies that beyond a certain point, further increases in $n$ do not significantly alter the transmission properties of the system. This behavior is graphically shown in Fig. \ref{sat_G}. It is seen from this figure that as $n$ increases, the transmission profile does not change significantly for large $n$.\\
\indent
We illustrate this further in Fig. \ref{sat_n} through density plots of $T$ over $n-k$ plane with relatively larger values of $n$. The range of $n$  is from 1 to 5 in Fig. \ref{sat_n}a, showcasing the initial variation in $T$ and the emergence of saturation. Fig. \ref{sat_n}b shows the extended range of \( n \) up to 10, where the pattern of \( T \) exhibits clear saturation, indicating minimal changes in transmission with increasing \( n \). Both the density plots illustrate that for larger values of \( n \), the variations in the transmission coefficient become less pronounced, indicating that changes in \( n \) do not significantly impact \( T \) after a certain point. This aligns with the theoretical expectations where the SVC(\( \rho, n \)) potential configuration would be less sensitive to the removal of smaller segments as \( n \) becomes large, which would in turn lead to a saturation of the transmission coefficient \( T \) with respect to \( n \). This behavior is likely due to the diminishing influence of the finer structural adjustments on the overall transmission characteristics of the potential system.
\section{Reflection coefficient: scaling feature}
\label{scaling}
\begin{figure}[H]
\begin{center}
\includegraphics[scale=0.55]{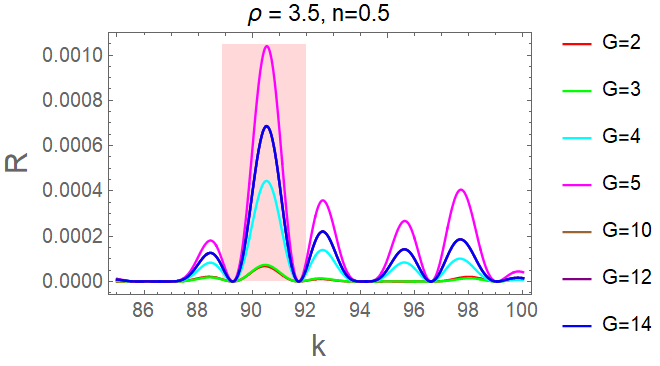} a \\
\includegraphics[scale=0.55]{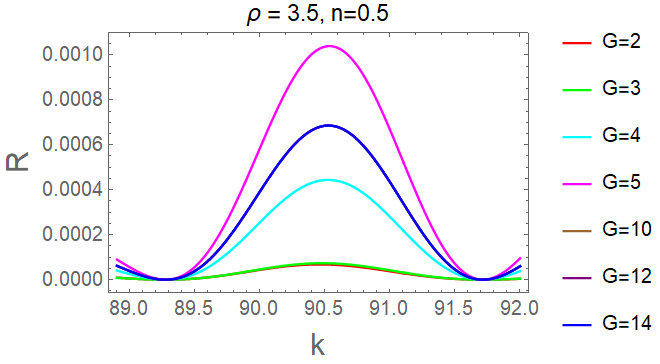} b 
\caption{\it Plot showing the variation of $R_{G}(k)$ for different stages $G$. Here, potential parameters are $L=2$ and $V_{0}=10$ and other parameters are mentioned in the plots. Fig. (b) is a magnified view of the highlighted region (light red hue) of Fig. (a). It is evident from the plots that the reflection profiles show significant differences in reflection profile for lower stages i.e. $G= 2, 3, 4, 5$, and invisible difference (complete overlap) for higher stages i.e. $G= 10, 12, 14$. This shows the convergence of the product term of Eq. (\ref{rG}).}
\label{convergence}
\end{center}
\end{figure}

In the limit of a significantly large value of $k$ the reflection coefficient $R(k)$ becomes highly attenuated. Under such conditions, it is possible to approximate the behavior of $R(k)$ using the expression provided in Eq. (\ref{T_svc}) as
\begin{equation}
    R(k) \sim 4^{G} \vert  M_{12} \vert ^{2} \prod_{i=1}^{G} \Omega_{i}^{2}.
    \label{r_small_value}
\end{equation}
When \( k \) is sufficiently large, the ratio \(\frac{V}{k^2}\) becomes very small, allowing for a first-order Taylor expansion. This results in an approximation for the magnitude squared of \( {M}_{12} \) as:
\begin{equation}
    \vert M_{12} \vert ^{2} \sim (Vl_{G})^{2}\frac{1}{4k ^{2}}. 
    \label{m12_small_value}
\end{equation}
Incorporating this into the expression for \( R(k) \), we arrive at a simplified form that highlights the inverse square dependence on \( k \):
\begin{equation}
    R(k) \sim 4^{G} \left( \frac{V l_{G}}{2} \right)^{2} \frac{1}{k^{2}} \prod_{i=1}^{G} \Omega_{i}^{2}.
\label{R_small_value}
\end{equation}
To maintain the total potential barrier area constant at each stage \( G \), we set the potential height \( V_{G} \) at each stage $G$ as:
\begin{equation}
    V_{G} = \frac{L}{2^{G} l_{G}} V_{0},
    \label{vg_values}
\end{equation}
where \( V_{0} \) is the initial potential barrier height (height of the potential for $G=0$) and \( l_{G} \) is the unit cell width of the rectangular barrier at stage $G$ defined by  Eq. (\ref{l_GG}). If we denote $R_{G}(k)$ as the reflection coefficient at each stage $G$, with each segment's potential height set at $V_{G}$, it can be demonstrated that the following relation holds (valid for large $k$),
\begin{equation}
    \frac{R_{G}(k)}{V_{0}^{2}} \sim \frac{L^{2}}{4k ^{2}} \prod_{i=1}^{G} \Omega_{i}^{2}.
    \label{rG}
\end{equation}
Eq. \ref{rG} shows that if potential height at stage $G$ is defined by Eq. \ref{vg_values}, then reflection probability will fall off as $\frac{1}{k^2}$ for large $k$. In Fig. \ref{convergence}, we illustrate the behavior of the function $R_{G}(k)$ as a function of the stage $G$ within the context of the SVC$(\rho, n)$ potential.

\begin{figure}[H]
\begin{center}
\includegraphics[scale=0.285]{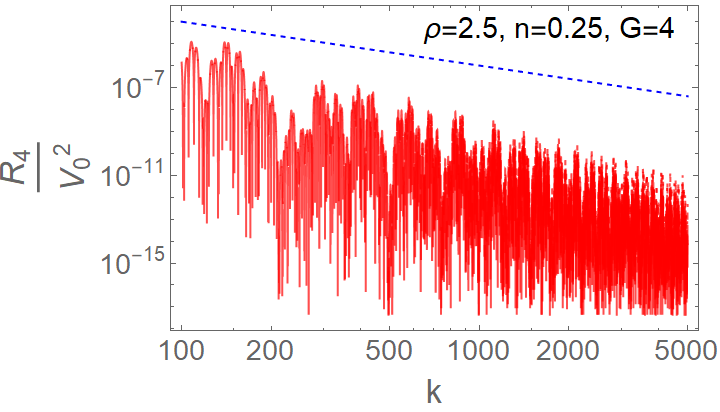} a \includegraphics[scale=0.285]{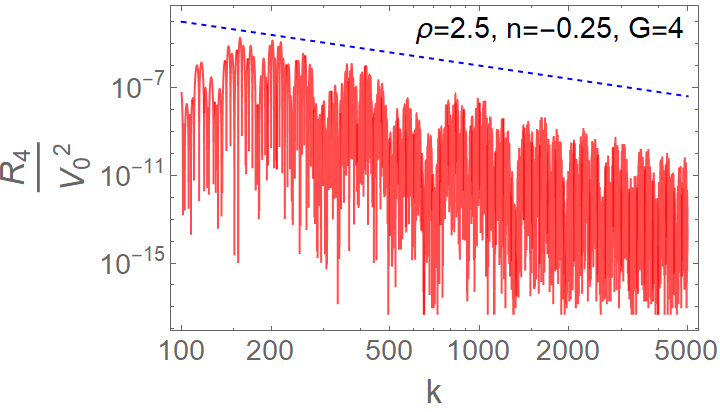}  b \\
\includegraphics[scale=0.285]{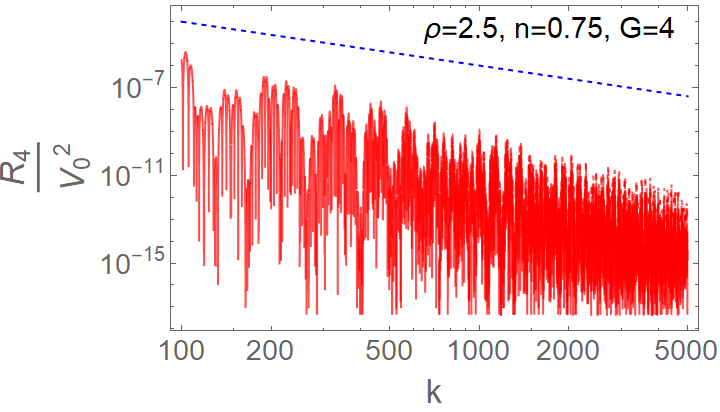} c \includegraphics[scale=0.285]{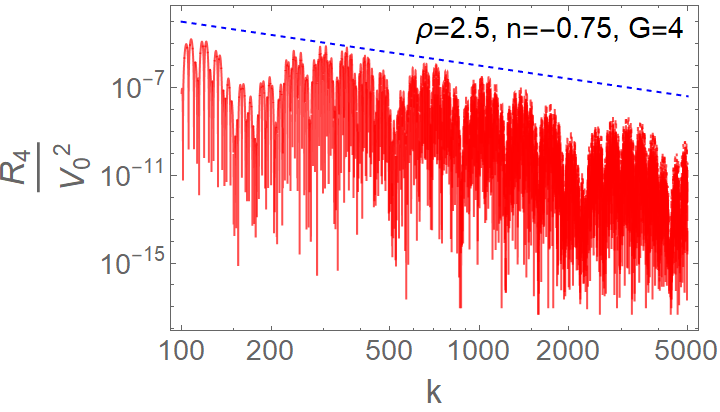}  d \\
\includegraphics[scale=0.285]{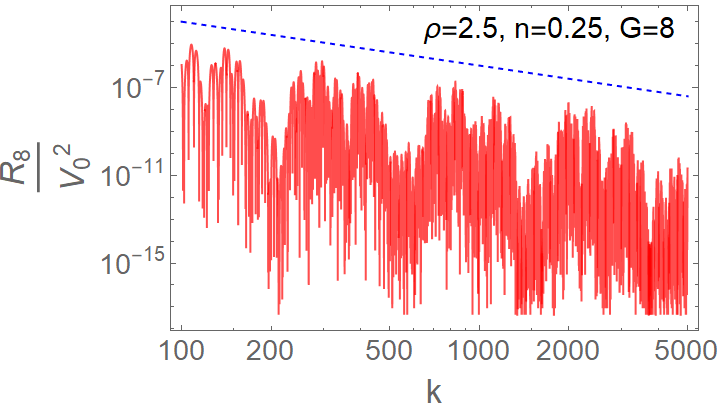} e  \includegraphics[scale=0.285]{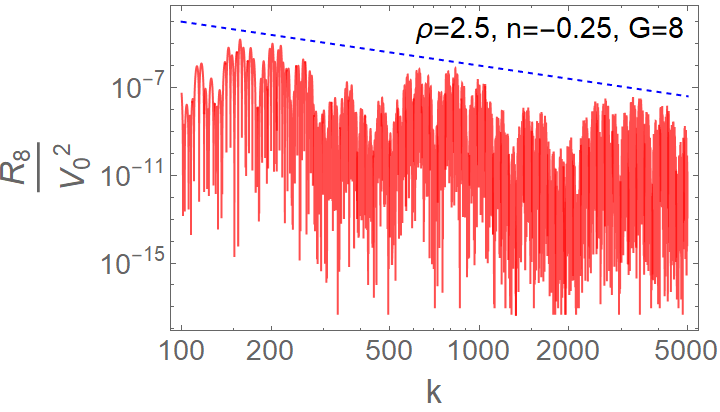}  f \\
\includegraphics[scale=0.285]{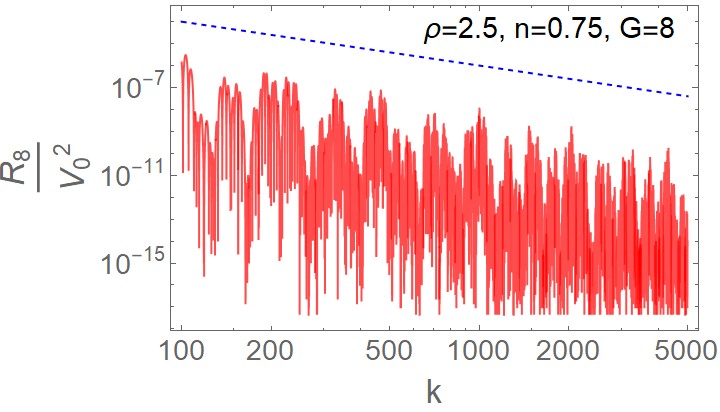} g  \includegraphics[scale=0.285]{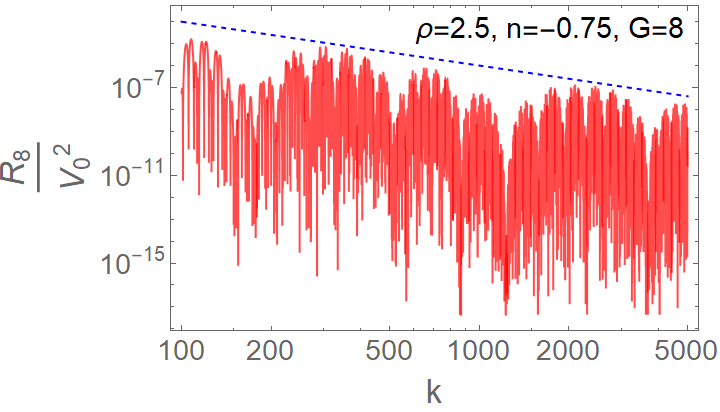}  h \\
\caption{\it Log-Log plots illustrating the scaling behavior of the reflection coefficients $\frac{R_4}{V_{0}^2}$ (displayed in the first and second rows) and $\frac{R_8}{V_{0}^2}$ (shown in the third and fourth rows) for large $k$ as $\frac{1}{k^{2}}$, represented by a dotted blue line. These plots are generated for two parameters: $\rho=2.5$ and $n$, with the potential parameters held constant at $L=1$ and $V_0=10$. The first column represents positive values of $n$, whereas the second column corresponds to negative values of $n$.}  
\label{scaling01}
\end{center}
\end{figure}

Notably, for higher stage $G$, the discernible differences in the profile of $R_{G}(k)$ diminish, indicative of the fast convergence of the product term described in Eq. (\ref{rG}). In Fig. \ref{convergence}b, we present a magnified representation of the prominently highlighted region (depicted in a light red hue) from Fig. \ref{convergence}a. This magnified view clearly reveals a congruence between the profiles of $R_{10}(k)$, $R_{12}(k)$, and $R_{14}(k)$, whereby they exhibit complete overlap. This trend aligns with prior findings observed in a previous study concerning the standard Cantor potential \cite{cantor_f7}, general SVC \cite{svc_tunneling} and unified Cantor potential \cite{ucp}.
Moreover, in light of the convergence properties inherent in the product term, particularly when evaluated at $V_{G}$ with increasing $G$, it is evident from the formulation in Eq. (\ref{rG}) that the behavior of $R_{G}(k)$ will exhibit a scaling relationship with $k$, characterized by an inverse square dependence for large values of $k$. This scaling behavior, akin to what has been previously established for the standard Cantor potential in reference \cite{cantor_f7}, is visually represented for the SVC$(\rho,n$) potential system in Fig. \ref{scaling01}.
\section{Conclusion and Discussions}
\label{conclusion}
In this research, we have unveiled the Smith-Volterra-Cantor (SVC) potential of power \( n \), a novel potential system that acts as a bridge between the well-studied general Cantor (GC) and the general Smith-Volterra-Cantor (GSVC or SVC(\(\rho\))) potentials. This SVC(\(\rho, n\)) potential not only extends the family of Cantor type potentials but also introduces a systematic approach to modulate between fractal and non-fractal characteristics by varying the power \( n \), thereby enriching the understanding of quantum mechanical systems.

Our analysis has been rooted in the super periodic potential (SPP) formalism, which facilitated the derivation of an analytical expression for the transmission coefficient \( T(k) \). The most striking revelation from our study is the emergence of exceedingly sharp transmission resonances within the SVC(\(\rho, n\)) potential system. These resonances are of particular interest because they suggest the potential's suitability for designing quantum filters with very narrow bandwidths.

The behavior of the SVC(\(\rho, n\)) potential is distinctly influenced by its two principal parameters, \( \rho \) and \( n \). We have meticulously shown that as the value of \( n \) increases, there is a saturation in the transmission profile, implying that changes in the potential structure have progressively less impact on the transmission properties. This saturation behavior is an indication of the potential's stability and could play a crucial role in applications where controlled quantum transmission is desired.

Further into our exploration, we have examined the reflection coefficient \( R_{G}(k) \), where we ensured that the total area under the potential barrier remained constant with each stage \( G \). The analyses present a compelling case that \( R_{G}(k) \) scales as \( \frac{1}{k^{2}} \) for large \( k \), which has been corroborated through both graphical and analytical methods. This scaling law has been observed for standard Cantor and  GSVC  potential systems and in the present work found to hold for  SVC(\(\rho, n\)) potential systems that unfies both these systems.

In conclusion, the SVC(\(\rho, n\))  potential's unique ability to blend fractal and non-fractal elements opens up new directions for theoretical and experimental research. Future studies could delve into the full implications of the new parameters and explore the broader class of potentials this research has hinted at potentially revolutionizing the way we understand and utilize quantum potentials obtained through the division of real line in a systematic way. A possible extension of the present work would be towards a broader class of potentials that can be realized by removing a \(\frac{1}{\rho^{a_{0}+ a_{1}G+ a_{2}G^{2}+\ldots+ a_{n}G^{n}}}\) portion from the middle of each segment from the segments of previous stages. This polynomial type Cantor system could unify a large class of potential systems and can be studied using the SPP formalism to investigate transmission characteristics. It will be interesting to explore such potential in future work.
\paragraph{}
{\it \bf{Acknowledgments}}:
The present investigation has been carried out under financial support from BHU- IoE scheme fellowships to VNS from Banaras Hindu University (BHU), Varanasi. MH acknowledges support from SPO-ISRO HQ for the encouragement of research activities. MU acknowledges the support from OPC Department, IIT Delhi for the encouragement of research activities. BPM acknowledges the support from the research grant under IoE scheme (Number - 6031), BHU, UGC Government of India.
\newpage
\bibliographystyle{elsarticle-num}
\bibliography{References}
\end{document}